\documentclass[12pt,journal,onecolumn]{IEEEtran}

\usepackage{cite,authblk,rotating,setspace,amsmath,amssymb,theorem,epstopdf}
\usepackage[caption=false, font=footnotesize]{subfig}

\newtheorem{theorem}{Theorem}
\newtheorem{lemma}{Lemma}
\newtheorem{corollary}{Corollary}

\newenvironment{Proof}[1]{\medskip\par\noindent{\bf Proof:\,}\,#1}{{\mbox{\,$\blacksquare$}\par}}

\allowdisplaybreaks

\begin{document}

\title{Age of Information in Multihop Multicast Networks\thanks{This work was supported by NSF Grants CNS 15-26608, CCF 17-13977 and ECCS 18-07348. This paper was presented in part at the Annual Asilomar Conference on Signals, Systems, and Computers, Pacific Grove, CA, October 2018.}}               

\author[1]{Baturalp Buyukates}
\author[2]{Alkan Soysal}
\author[1]{Sennur Ulukus}
\affil[1]{\normalsize Department of Electrical and Computer Engineering, University of Maryland, MD}
\affil[2]{\normalsize Department of Electrical and Electronics Engineering, Bahcesehir University, Istanbul, Turkey}

 %\normalsize {\it ypwei@umd.edu  \qquad \it kbanawan@umd.edu} \qquad {\it ulukus@umd.edu}}         

\maketitle

\vspace*{-1.0cm} 

\begin{abstract}
	We consider the age of information in a multihop multicast network where there is a single source node sending time-sensitive updates to $n^L$ end nodes, and $L$ denotes the number of hops. In the first hop, the source node sends updates to $n$ first-hop receiver nodes, and in the second hop each first-hop receiver node relays the update packets that it has received to $n$ further users that are connected to it. This network architecture continues in further hops such that each receiver node in hop $\ell$ is connected to $n$ further receiver nodes in hop $\ell+1$. We study the age of information experienced by the end nodes, and in particular, its scaling as a function of $n$. We show that, using an earliest $k$ transmission scheme in each hop, the age of information at the end nodes can be made a constant independent of $n$. In particular, the source node transmits each update packet to the earliest $k_1$ of the $n$ first-hop nodes, and each first-hop node that receives the update relays it to the earliest $k_2$ out of $n$ second-hop nodes that are connected to it and so on. We determine the optimum $k_\ell$ stopping value for each hop $\ell$ for arbitrary shifted exponential link delays.	
\end{abstract}

\section{Introduction}
Recently, with the increase in the number of communication network applications requiring real-time status information, timeliness of the received messages has become a critical and desirable feature for networks. Such applications include sensor networks measuring ambient temperature \cite{Mainwaring02}, autonomous vehicular networks where instantaneous vehicle information including velocity, position and acceleration is needed \cite{Papadimitratos09} and news reports from Twitter. In all these applications, information loses its value as it becomes stale.

This motivates the study of age of information, which is a metric measuring the freshness of the received information. A typical model to study age of information includes a source which acquires time-stamped status updates from a physical phenomenon. These updates are transmitted over the network to the receiver(s) and the age of information in this network, or simply the age, is the time elapsed since the most recent update at the receiver was generated at the transmitter. In other words, at time $t$, age $\Delta(t)$ of a packet which was generated at time $u(t)$ is $\Delta(t)=t-u(t)$.

Most of the existing work focuses on age analysis in a queuing-theoretic setting. References \cite{Kaul12a, Costa16, Yates12} study the age under various arrival and service profiles. References \cite{Najm17} and \cite{Soysal18} investigate packet management strategies including blocking and preemption for M/G/1/1 and G/G/1/1 queues, respectively. Reference \cite{Bedewy17a} studies multi-hop networks in which update packets are relayed from one node to another. Another line of research studies the age under the energy harvesting setting \cite{Arafa17b, Arafa17a, Yates15, Bacinoglu15, Wu18, Arafa18a, Arafa18b, Arafa18d, Baknina18a, Baknina18b}.

Considering dense IoT deployments and the increase in the number of users in networks supplying time-sensitive information, the scalability of age as a function of the number of nodes has become a critical issue. To this end, we need to study how the age performance of the network changes with growing network size. Reference \cite{Ioannidis09} studies a mobile social network with a single service provider and $n$ communicating users, and shows that under Poisson contact processes among users and uniform rate allocation from the service provider, the average age of the content at the users grows logarithmically in $n$. In contrast, reference \cite{Zhong17a} observes that in a single-hop multicast network appropriate stopping threshold $k$ can prevent information staleness as the network grows. 

\begin{figure}[t]
	\centering  
	\includegraphics[width=.6\columnwidth]{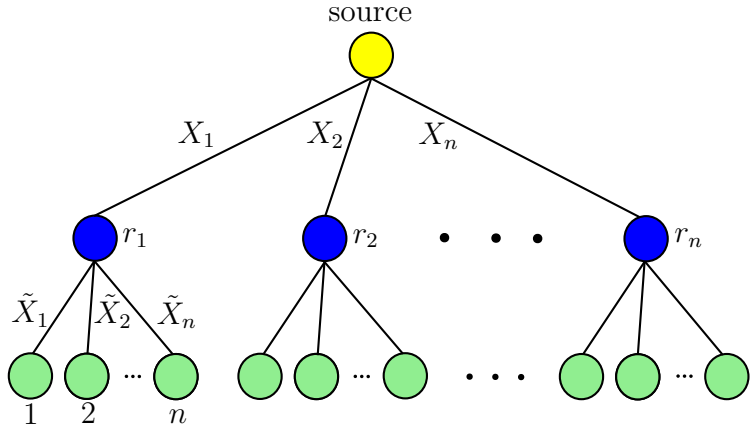}
	\caption{Two-hop multicast network with a single source node sending updates through $n$ middle nodes each of which is tied to further $n$ end nodes.}
	\label{fig:model}
\end{figure}

Motivated by this observation, we study the scalability of the age in multihop multicast networks using similar threshold ideas. Extending the results of \cite{Zhong17a}, we first analyze the single-hop problem with exogenous arrivals where the source directly communicates with the end users but cannot generate the updates itself. We then characterize the age for the two-hop case (see Fig.~\ref{fig:model}) using our single-hop with exogenous arrivals result as a building block. We show that for this two-hop multicast network under i.i.d.~shifted exponential link delays and stopping thresholds $k_1$ and $k_2$ at each hop, an upper bound on the average age can be obtained. Through this upper bound, we show that the average age is limited by a constant as $n$ increases. Utilizing this upper bound, we then extend the result of two-hop network to $L$-hop multicast networks for general $L$. We determine the optimal stopping threshold for each hop $\ell$, $k_\ell$, that minimizes the average age for arbitrary shifted exponential link delays.

\section{System Model and Age Metric} \label{model}
We consider a multihop system where in the first hop a single source node broadcasts time-stamped updates to $n$ first-hop receiver nodes using $n$ links with i.i.d.~random delays, and in the second hop, each first-hop receiver node relays the update packets it has received to $n$ further nodes that are connected to it. This network architecture continues in further hops such that each receiver node in hop $\ell-1$ acts as a transmitter in hop $\ell$ and it is connected to $n$ further receiver nodes in hop $\ell$. Fig.~\ref{fig:model} shows the described model for two hops. We first consider the age of information in a two-hop ($L=2$) multicast network and then extend our results to general multihop multicast networks with $L$ hops, where $L$ is arbitrary. 

We start describing our system model within the simpler two-hop setting. In the two-hop multicast network, an update takes $X$ time to reach from the source node to a particular mid-level node and $\tilde{X}$ time to reach from the mid-level nodes to a particular end node where $X$ and $\tilde{X}$ are shifted exponential random variables with parameters $(\lambda,c)$ and $(\tilde{\lambda},\tilde{c})$, respectively, where $c$ and $\tilde{c}$ are positive constants. When the shift parameters are zero random variables corresponding to service times become exponentially distributed. In each hop, the service times of individual links are i.i.d. realizations of random variables $X$ and $\tilde{X}$, e.g., in the first hop service times of individual links are i.i.d. $X_i$ and in the second hop service times of individual links are i.i.d. $\tilde{X}_i$.

In the two-hop scenario, age is measured for each of the $n^2$ end nodes and for node $i$ at time $t$ age is the random process $\Delta_i(t) = t - u_i(t)$ where $u_i(t)$ is the time-stamp of the most recent update at that node. When the source node sends out update $j$, it waits for the acknowledgment from the earliest $k_1$ of $n$ middle nodes. After it receives all $k_1$ acknowledgement signals, we say that update $j$ has been completed and the source node generates update $j+1$. At this time, transmissions of the remaining $n-k_1$ packets are terminated. Thus, if a node in the first hop is not in the earliest $k_1$ nodes to receive update $j$ then service of this update is preempted. In the second hop, these earliest $k_1$ nodes that have received update $j$ start transmitting this update to their end nodes and they stop whenever $k_2$ of their end nodes have received the current update.  Middle nodes implement a blocking scheme when they are busy transmitting to the end nodes, i.e., they discard arriving packets when they are not idle.  When the middle nodes finish transmitting the current update to $k_2$ of their children nodes, they wait for the arrival of the next update from the source node.
	
\begin{figure}[t]
	\centering
	\subfloat[\label{}]{\includegraphics[width=.5\columnwidth]{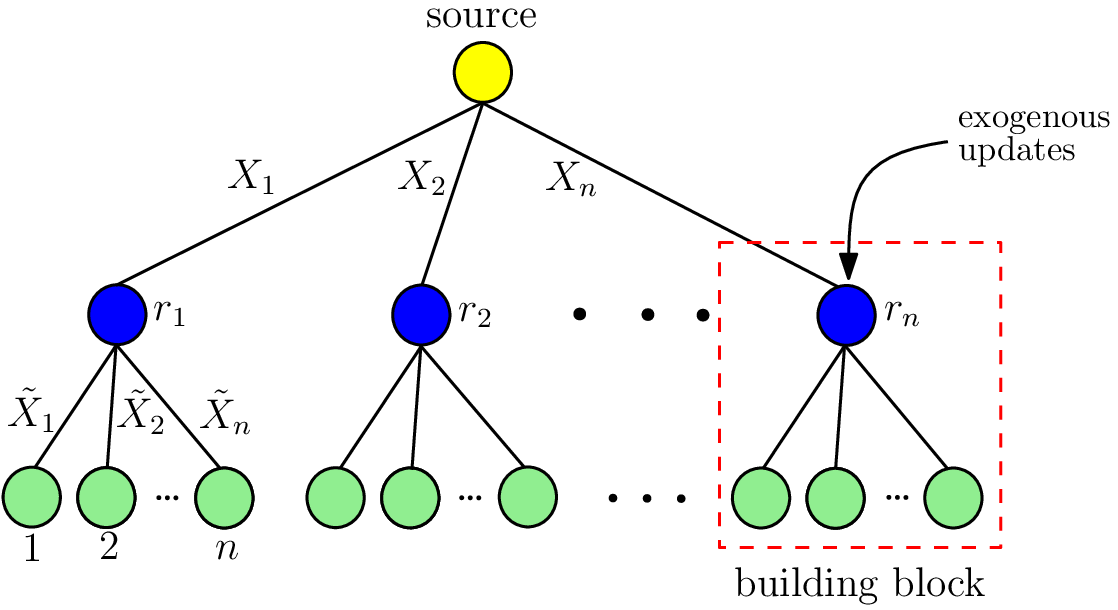} }%
	\subfloat[\label{}]{\includegraphics[width=.5\columnwidth]{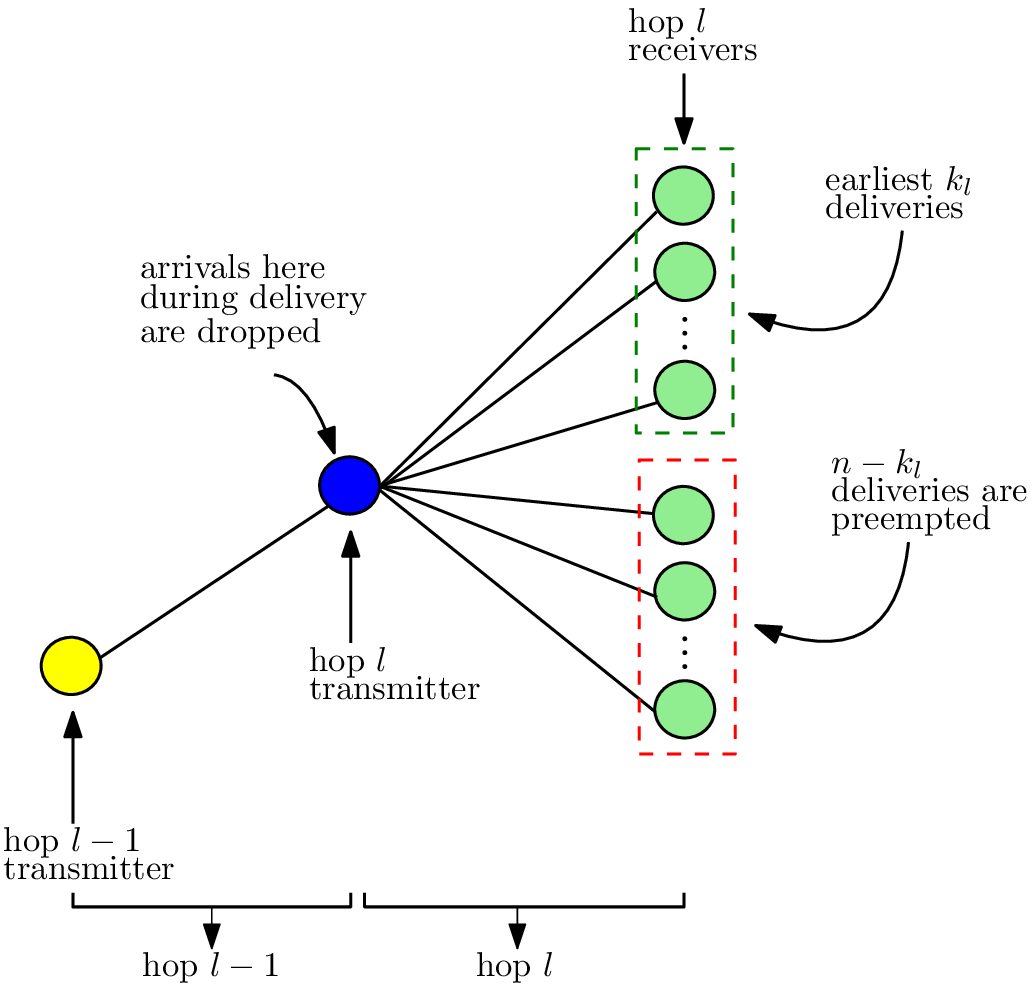}}
	\caption{Multicast network model: (a) two-hop operation, (b) details of hops $\ell-1$ and $\ell$.}
	\label{fig:1.5}
\end{figure}	

For general $L$, principles that are explained above are repeated at every hop, e.g., the transmitters in hop $\ell$ wait for $k_\ell$ of their children nodes to receive the current update before they declare that the current update has been completed; they drop all other incoming updates from transmitters in hop $\ell-1$ as they are transmitting the current update; and when their update is received by $k_\ell$ receivers they preempt the remaining $n-k_\ell$ updates. When this is over, transmitters in hop $\ell$ wait for the next update from their parent nodes in hop $\ell-1$; see Fig.~\ref{fig:1.5}(b).

The metric we use, time averaged age, is given by
\begin{align}
\Delta = \lim_{\tau\to\infty} \frac{1}{\tau} \int_{0}^{\tau} \Delta(t) dt
\end{align}
where $\Delta(t)$ is the instantaneous age of the last successfully received update as defined above. We will use a graphical argument similar to \cite{Zhong17a} to derive the average age at an individual end node. Since all link delays are i.i.d. for all nodes and packets, each end node $i$ experiences statistically identical age processes and will have the same average age. Therefore, it suffices to focus on a single end user for the age analysis.

\section{Building Block: Single-Hop Network with Exogenous Arrivals} \label{section:building-block}

\begin{figure}[t]
	\centering  
	\includegraphics[width=.5\columnwidth]{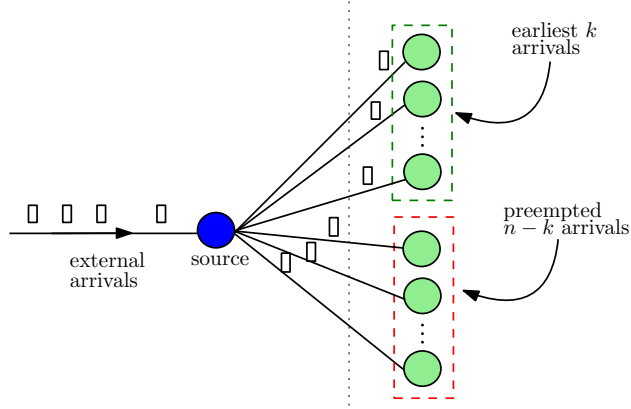}
	\caption{The arrival and update structure of a building block.}
	\label{fig:1.6}
\end{figure}

We first note that at the second hop what we essentially have is $n$ parent nodes each tied to $n$ children nodes. Therefore, each second hop transmitter and its children nodes correspond to the single-hop network analyzed in \cite{Zhong17a} with one important difference: second hop transmitters cannot generate update packets. They can only relay packets sent from the source node. Thus, in this section, we first analyze a single-hop network in which update packets arrive exogenously with a given expected interarrival time $\frac{1}{\mu}$. Then, using this network as a building block we analyze the $L$-hop network described in Section~\ref{model}, first for $L=2$, i.e., a two-hop network, then for general $L$.  We have i.i.d.~shifted exponential service times between the source and each of its $n$ children nodes as in \cite{Zhong17a}. Similarly, transmission of the current update stops when $k$ out of $n$ nodes receive the update. Thus, in this section, we extend the results of \cite{Zhong17a} to the case of exogenous arrivals, and determine a $k$ threshold which depends on $\lambda$, $c$ and $\mu$. Fig.~\ref{fig:1.6} shows the arrival and update structure of a building block. Fig.~\ref{fig:1.5}(a) shows this building block as part of a two-hop network; and Fig.~\ref{fig:1.5}(b) shows it as a part of an $\ell$th hop in an $L$-hop network. 

Under this model with i.i.d. link delay $X$, an update takes $X_{k:n}$ units of time to reach $k$ out of $n$ nodes where we denote the $k$th order statistic of random variables $X_1, \ldots ,X_n$ as $X_{k:n}$. Here $X_{k:n}$ is the $k$th smallest of $X_1, \ldots ,X_n$, e.g., $X_{1:n} = \min\{X_i\}$ and $X_{n:n} = \max\{X_i\}$. For shifted exponential random variable $X$, we have
\begin{align}
E[X_{k:n}] =& c + \frac{1}{\lambda}(H_n - H_{n-k}) \\
Var[X_{k:n}] =& \frac{1}{\lambda^2}(G_{n} - G_{n-k})
\end{align}
where $H_n = \sum_{j=1}^{n} \frac{1}{j}$ and $G_{n} = \sum_{j=1}^{n} \frac{1}{j^2}$. Using these,
\begin{align}
E[X_{k:n}^2] =& c^2 + \frac{2c}{\lambda}(H_n - H_{n-k}) + \frac{1}{\lambda^2}\left((H_n - H_{n-k})^2 +G_{n} - G_{n-k} \right)
\end{align}

\begin{figure}[t]
	\centering  \includegraphics[width=.8\columnwidth]{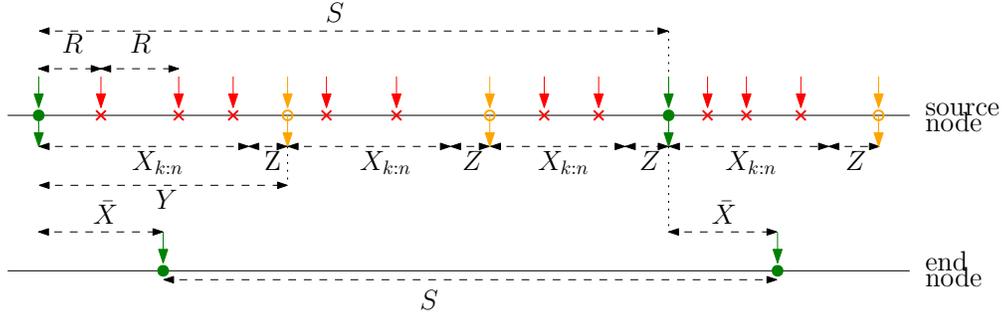}
	\caption{Update process for an end node in the building block setting.  Filled circles indicate the arrivals successfully received by this particular end node. Crosses indicate updates that arrive at the source node when it is busy transmitting another update. Empty circles show the updates that are transmitted by the source node but are not received by this particular end node, i.e., they are preempted from the perspective of this particular end node.}
	\label{fig:updates}
\end{figure}

We say that the system is busy when a transmitted update has not been received by $k$ out of $n$ nodes yet. Once $k$ end nodes receive an update, transmitter stops the service to the remaining $n-k$ nodes. Then, the system is idle until the next update arrives. Fig. \ref{fig:updates} shows a realization of the update process between the source node and a particular end node. Note that realizations of the update process for different end nodes might be different depending on when and for which updates these nodes have been one of the earliest $k$ nodes. On the other hand, although realizations are different, each end node experiences the same random process. Updates arrive at the source node with an interarrival time, $R$, where $E[R]=\frac{1}{\mu}$. In the most general setting, $R$ is an arbitrary i.i.d. random variable. In Fig.~\ref{fig:updates}, arrows that are above and below the source node line indicate the received and transmitted updates at the source node, respectively. 

An important aspect of our model is that updates at the source node are divided into three groups, namely successful updates, dropped updates, and preempted updates. In Fig.~\ref{fig:updates}, filled circles correspond to successfully received updates. They indicate that the update received at the source node started transmission to $n$ end nodes and this particular end node received the update. We denote the time between the transmission and the reception of successful updates (filled circles) as the service time, $\bar X$. The order of this particular node might be smaller than $k$, thus $\bar X \leq X_{k:n}$. The source continues the service of the first filled circled update until $k$ nodes receive the update. During this time if new updates arrive at the source, they are dropped; the dropped updates (crosses in Fig.~\ref{fig:updates}) never go into service, they are lost. Once $k$ nodes receive this update, system becomes idle and the source waits for the arrival of the next update. 

We denote the waiting time until the next arrival with $Z$. The next arrival which is shown with an empty circle in Fig.~\ref{fig:updates} starts the service. However, for this particular realization of the update process, this particular end node is not one of the earliest $k$ end nodes during the transmission of empty circled update. Since the source stops service once $k$ end nodes receive the update, this transmitted update never arrives at this particular end node. We denote those updates that start a service but do not arrive at this particular end node as preempted updates. Even though preemption usually means stopping a current service in order to start a new one immediately, here we use the word preemption to mean that the current service is stopped, and a new service will start after an idle period when a new update arrives at the source. However, similar to regular meaning of preemption, in our model as well, current preempted update leaves service and a fresh update takes over. 

We denote the time between two consecutive departures from the source as $Y = X_{k:n} + Z$. Note that there are a random number of dropped updates during each realization of $Y$. In Fig.~\ref{fig:updates}, between the first successful (filled circle) update and the next update that starts the service (empty circle), there are three dropped updates (crossed). In other words, the fourth received update is able to start a new service. In addition, there are multiple realizations of $Y$ before the next successful (filled circle) update, since this particular end node is not able to receive empty circled updates. We denote the time between two successful updates, i.e., two consecutive updates that depart from the source and successfully arrive at the end node, with $S$. Remember that $Y$ is the time between two updates that depart from the source and start the service but are not guaranteed to arrive at this particular end node. We can relate the interarrival times of departing updates and arriving updates using $S = \sum_{i=1}^{M} Y_i $. In Fig. \ref{fig:updates}, this particular end node receives the first and the fourth updates that depart the source and enter the service. Therefore, in this example we have $M=3$. In addition, remember that $R$ is the interarrival time between the updates that arrive at the source node. For a general $R$, there is no closed-form expression with known variables that relates interarrival times at the end node, $S$, to interarrival times at the source node, $R$. However, when $R$ is exponential $Z$ is exponential as well due to the memoryless property of the exponential distribution. Thus, for exponential $R$ we have $S = \sum_{i=1}^{M} (X_{k:n}+R)_i$.

When the current update reaches $k$ earliest nodes, the source node terminates the remaining $n-k$ transmissions and begins to wait for the next arrival and then repeats the process when the next update packet arrives. Since the link delays are i.i.d., the end users receive the packet in service with probability $p = \frac{k}{n}$. If an end user receives update $j$ and the next one it receives is update $j+M$, then $M$ is geometrically distributed with $p$ with moments
\begin{align}
E[M] = \frac{1}{p} = \frac{n}{k}, \qquad 
E[M^2] = \frac{2-p}{p^2} = \frac{2n^2}{k^2}-\frac{n}{k} \label{moment4}
\end{align}

Similar to \cite{Zhong17a}, the average age for the earliest $k$ stopping scheme with exogenous packet arrivals with rate $\mu$ is
\begin{figure}[t]
	\centering  \includegraphics[width=0.5\columnwidth]{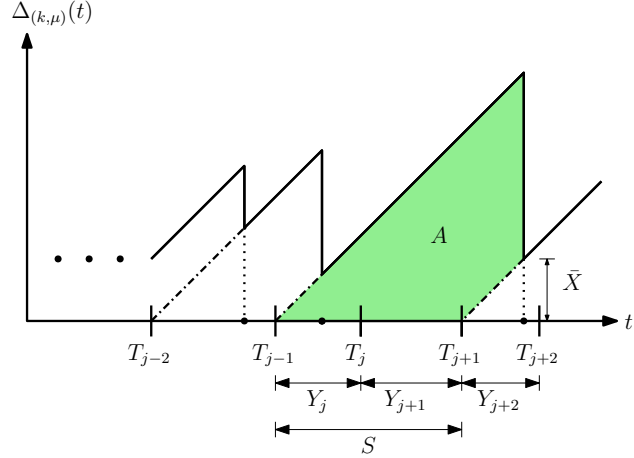}
	\caption{Sample age evolution $\Delta_{(k,\mu)}(t)$ of an end node. Updates that find the system idle arrive at times $T_i$ at the source. Here, update $j$ arrives at time $T_{j-1}$ and immediately goes into service. Successful update deliveries are marked with $\bullet$ and in this figure, updates $j-1$, $j$ and $j+2$ are delivered successfully whereas update $j+1$ is terminated.}
	\label{fig:ageEvol}
\end{figure}
\begin{align}
\Delta_{(k,\mu)} =& \frac{E[A]}{E[S]}  \label{age_kmu}
\end{align}
where $A$ denotes the shaded area in Fig.~\ref{fig:ageEvol} and $S$ is its length. Remember from Fig. \ref{fig:updates} that the random variable $S$ is the  interarrival time at the end nodes. Inspecting Fig.~\ref{fig:ageEvol} to calculate $A$, we find $E[A] = \frac{1}{2}E[S^2] + E[S]E[\bar X]$. Here, $\bar{X}$ denotes the service time of a successful update such that $E[\bar{X}] = E[X_i | i \in \mathcal{K}]$ where $\mathcal{K}$ is the set of earliest $k$ nodes that receive the update. Now, (\ref{age_kmu}) becomes\footnote{Since the building block studied in this section will eventually be used in the upcoming sections in the analysis of two-hop and $L$-hop networks, we should emphasize that the definitions of $\bar X$ and $S$ do not assume any network structure. As far as (\ref{eqn:age_kmu_add}) is considered, it is not important where an update that is received by an end node is generated. $\bar{X}$ is the time that the update spends in the system from the time it is generated (possibly by a node other than the one that relays it to the end node) until the time it is received at the end node. $S$ is the time between two consecutive successfully received updates, regardless of whether the updates are generated at the node that transmits them or not. This reasoning is important for our derivations in multihop multicast networks.}
\begin{align}
\Delta_{(k,\mu)} =& E[\bar{X}] + \frac{E[S^2]}{2E[S]}  \label{eqn:age_kmu_add}
\end{align}
 We can write the first and second moments of $S$  in terms of $Y$ as 
\begin{align}
E[S] &= E[M]E[Y] \\
E[S^2] &= E[M]E[Y^2]+E[Y]^2E[M^2-M]
\end{align} 
Inserting these into (\ref{eqn:age_kmu_add}) we obtain
\begin{align}
\Delta_{(k, \mu)} = E[\bar{X}]+ \frac{E[M^2]}{2E[M]}E[Y] + \frac{Var[Y]}{2E[Y]} \label{eqn:age_kmu2}
\end{align}

In the following theorem, we determine the age of an update at an end node for a single-hop building block model using (\ref{eqn:age_kmu2}).

\begin{theorem} \label{thm_kmu}
For  a single-hop building block model with exogenous arrivals that have expected interarrival time $\frac{1}{\mu}$, for the earliest $k$ stopping scheme, the average age of an update at an individual end node is
	\begin{align}
	\Delta_{(k,\mu)} =& \frac{1}{k} \sum_{i=1}^{k} E[X_{i:n}] + \frac{2n-k}{2k}(E[X_{k:n}] + E[Z]) + \frac{Var[X_{k:n}+Z]}{2(E[X_{k:n}] + E[Z])} \label{thm_kmu_res}
	\end{align}
\end{theorem}

\begin{Proof}
	The first term comes from $E[\bar{X}]$ as
	\begin{align}
	E[\bar{X}] =& E[X_j| j \in \mathcal{K}] = \sum_{i=1}^{k} E[X_{i:n}] Pr[j=i| j \in \mathcal{K} ] = \frac{1}{k} \sum_{i=1}^{k} E[X_{i:n}]
	\end{align}	
	where we used the fact that, since we have $k$ out of $n$ nodes selected independently and identically in $\mathcal{K}$, we have $Pr[j=i| j \in \mathcal{K}] = \frac{1}{k}$. The second and third terms are obtained upon substitution of $Y = X_{k:n} + Z$ and $E[M]$ and $E[M^2]$ expressions given in (\ref{moment4}) into (\ref{eqn:age_kmu2}).  
\end{Proof}

When we have general interarrival times as we have in this problem, $X_{k:n}$ and $Z$ may be dependent. However, with exponential interarrivals we can show their independence using the memoryless property, and simplify the age expression in (\ref{thm_kmu_res}) as follows.

\begin{corollary} \label{corr1}
	When the arrival process is Poisson with rate $\mu$, the age of an end node is
	\begin{align}
	\Delta_{(k,\mu)} =& \frac{1}{k} \sum_{i=1}^{k} E[X_{i:n}] + \frac{2n-k}{2k\mu}(\mu E[X_{k:n}] + 1) + \frac{\mu Var[X_{k:n}]}{2(\mu E[X_{k:n}] + 1)} + \frac{1}{2(\mu^2 E[X_{k:n}] + \mu)} \label{age_Poisson}
	\end{align}
\end{corollary}

\begin{Proof}
	When the arrival process is Poisson with $\mu$, in other words, $R$ is exponential with expected interarrival time $\frac{1}{\mu}$, the random variable $Z$ which corresponds to the residual interarrival time is exponentially distributed with the same parameter, i.e., $Z=R$. In addition, $Z$ is independent of $X$ due to the memoryless property of the exponential distribution. Then,
	\begin{align}
	Var[Y] = Var[X_{k:n}+Z] = Var[X_{k:n}]+Var[Z]
	\end{align}
	and we plug in $E[Z] = \frac{1}{\mu}$ and $Var[Z] = \frac{1}{\mu^2}$. 
\end{Proof}
When the service times $X$ are i.i.d. shifted exponential random variables and when $n$ is large, we can further simplify the age expression in (\ref{age_Poisson}) as follows.
\begin{corollary}  \label{corr2}
	For large $n$ and $n>k$, set $k = \alpha n$. For shifted exponential $(\lambda,c)$ service times $X$, the average age for the earliest $k$ scheme with exogenous Poisson arrivals with rate $\mu$ can be approximated as
	\begin{align}
	\Delta_{(k,\mu)} \approx \frac{c}{\alpha} + \frac{c}{2}  + \frac{1}{\lambda} - \frac{1}{2\lambda}\log(1-\alpha) +\frac{1}{\alpha \mu}-\frac{1}{2\mu} + \frac{1}{2}\left(\mu^2c \!- \frac{\mu^2 \log(1\!-\alpha)}{\lambda} \!+\mu\right)^{-1} \label{corr2_res}
	\end{align}	
\end{corollary}
\begin{Proof}
	Using the order statistics above,
	\begin{align}
	\delta_1 = & \frac{1}{k} \sum_{i=1}^{k} E[X_{i:n}] =  c + \frac{H_n}{\lambda} - \frac{1}{k\lambda}\sum_{i=1}^{k} H_{n-i} 
	\end{align}	 	
	As in \cite{Zhong17a}, we have $\sum_{i=1}^{k} H_{n-i}  = \sum_{i=1}^{n-1} H_i - \sum_{i=1}^{n-k-1} H_i$ and the series identity $\sum_{i=1}^{k} H_i  = (k+1)(H_{k+1}-1)$. Using these we get
	\begin{align}
	\delta_1 =  c\!+ \frac{1}{\lambda} \!- \!\frac{n\!-k}{k\lambda }(H_n \!- H_{n-k}) \! \approx c\!+\! \frac{1}{\lambda} \!+\! \frac{1\!-\!\alpha}{\alpha\lambda }\!\log(1\!-\!\alpha)
	\end{align}
	since for large $n$, we have $H_i \approx \log(i)+\gamma$. Also,
	\begin{align}
	\delta_2 =& \frac{2n-k}{2\mu k} (\mu E[X_{k:n} ] + 1) \\ =& \frac{2n-k}{2\mu k}  \left(\mu \left(c+ \frac{H_n - H_{n-k}}{\lambda}\right)+1\right) \\
	\approx& \frac{(2-\alpha)c}{2\alpha} + \frac{\alpha-2}{2\alpha\lambda}\log(1-\alpha) + \frac{2-\alpha}{2\alpha\mu}
	\end{align}	
	Next, we note that we have 	
	\begin{align}
	\lim_{n\to\infty} \frac{\mu Var[X_{k:n}]}{2(\mu E[X_{k:n}]+1)} = 0 \label{lim_exp}
	\end{align}	 	
	We see this from the expected values of order statistics,
	\begin{align}
	\frac{\mu Var[X_{k:n}]}{2(\mu E[X_{k:n}]+1)} =& \frac{\mu (G_{n}-G_{n-k})}{2(\mu \lambda^2c+\mu \lambda(H_n-H_{n-k})+\lambda^2)} \label{var}
	\end{align}
	We know that the sequence $G_{n}$ converges to $\frac{\pi^2}{6}$. As $n$ increases $G_{n-k} = G_{(1-\alpha)n}$ also goes to the same value making the numerator 0. Thus, as $n$ tends to $\infty$ (\ref{lim_exp}) is achieved. Similarly,
	\begin{align}
	\delta_3 =& \frac{1}{2(\mu^2\!E[X_{k:n}]\!+\mu)}\approx \frac{1}{2}\!\left(\!\mu^2c\!-\frac{\mu^2 \log(1\!-\alpha)}{\lambda} \!+\mu\!\right)^{-1}
	\end{align}	 	
	Summing $\delta_1$, $\delta_2$, and $\delta_3$ yields the expression in (\ref{corr2_res}).
\end{Proof}

Note that the age expression in (\ref{corr2_res}) when $n$ is large is a function of the ratio $\alpha=\frac{k}{n}$ only implying that the age converges to a constant even when the packets arrive exogenously, similar to \cite{Zhong17a} where the packets are generated at will at the source. 

\begin{figure}
	\centering
	\subfloat[\label{fig:d1}]{\includegraphics[width=.5\columnwidth]{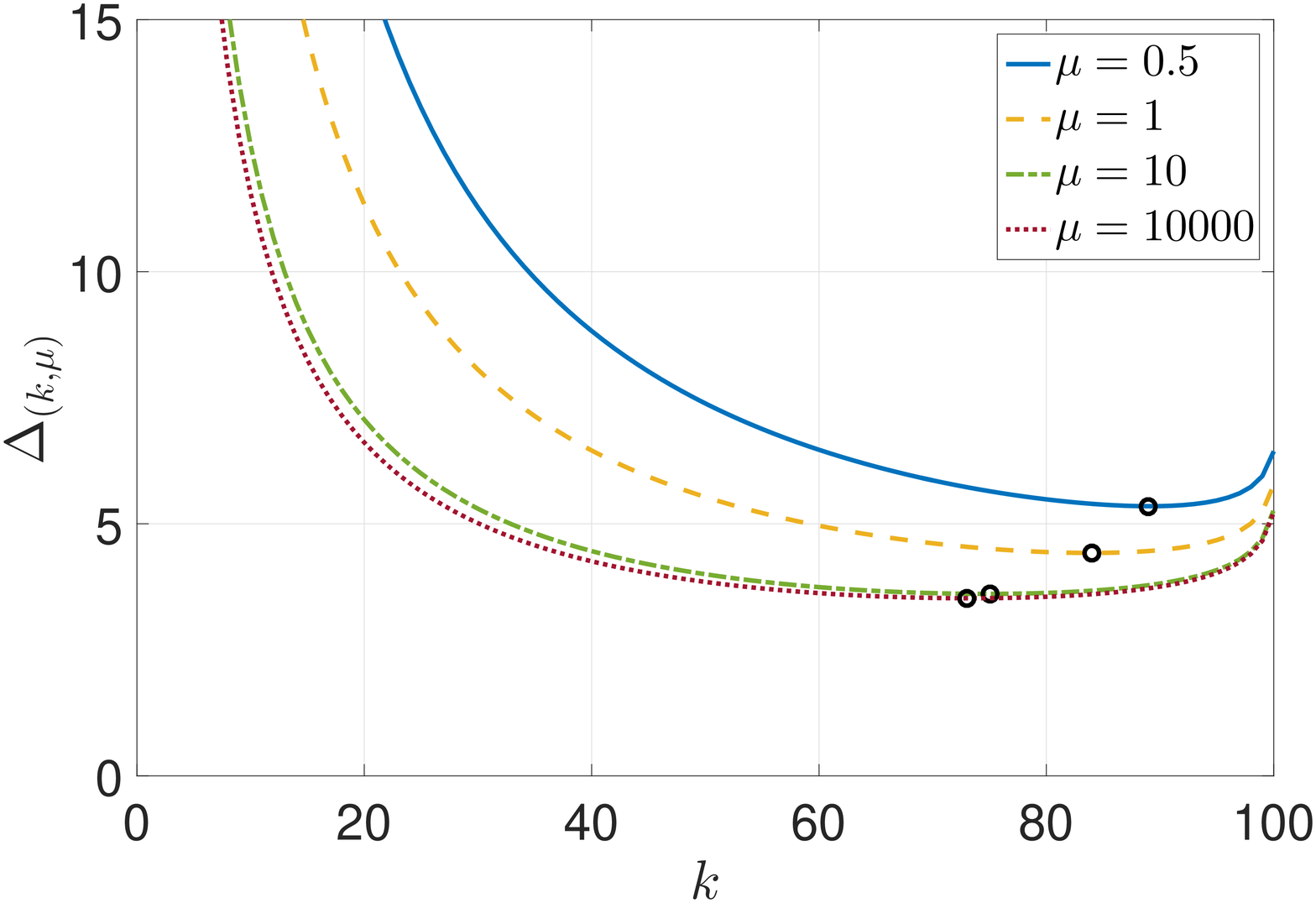} }%
	\subfloat[\label{fig:d2}]{\includegraphics[width=.5\columnwidth]{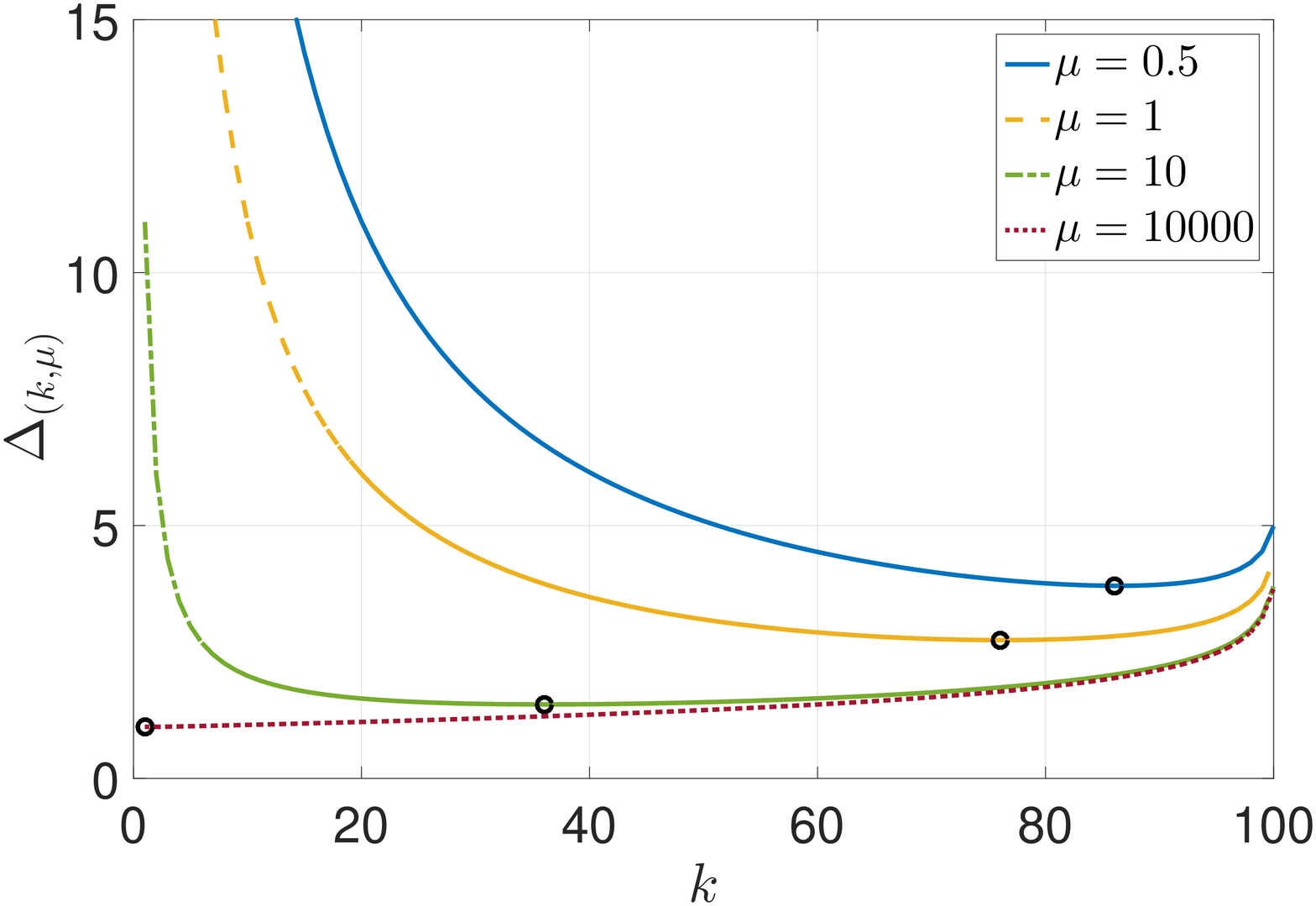} }
	\caption{$\Delta_{(k, \mu)}$ as a function of stopping threshold $k$ for several $\mu$ values with $\lambda =1$. $\circ$ marks the minimized $\Delta_{(k, \mu)}$: (a) when $c=1$, (b) when $c=0$. }
	\label{fig:buildblock}
\end{figure}

Although there is no explicit closed form solution for the optimal $\alpha$, denoted as $\alpha^*$, which minimizes (\ref{corr2_res}), we can calculate it numerically. For instance, when $(\lambda, c)=(1, 1)$ and Poisson arrival rate is $\mu=1$, age minimizing $\alpha$ is $\alpha^* = 0.837$. This optimal value is higher than that of the original case in which the source itself generates the packets at will, which is $\alpha^*=0.732$ found in \cite{Zhong17a}. This is because with exogenous arrivals at the source node, update interarrival time at the end nodes is larger. In this regard, we see that when the Poisson arrival rate is increased $\alpha^*$ decreases. This is expected because when the arrival rate is high (i.e., update packets arrive frequently), the source node prefers to wait for the \emph{freshest} one instead of sending the current update to more and more end users. Similarly, when the arrival rate is low (i.e., update packets arrive infrequently), $\alpha^*$ is higher because in this case source knows that interarrival time is higher so that before it waits for the next packet it wants to update as many end nodes as it can (see Fig.~\ref{fig:buildblock}(a)). 
We also note that as we take $\mu \to\infty$ in (\ref{corr2_res}), we get
\begin{align}
\Delta_{(k)} \approx \frac{c}{\alpha} + \frac{c}{2}  + \frac{1}{\lambda} - \frac{1}{2\lambda}\log(1-\alpha)
\end{align}	
which is the age expression in \cite{Zhong17a}. Thus, age expression under exogenous Poisson arrivals with rate $\mu$ converges to the case in which source generates the packets itself as $\mu$ tends to $\infty$.

Finally, we note an interesting aspect of the problem with exogenous arrivals: It is shown in \cite{Zhong17a} that when the service time random variable $X$ is exponential (i.e., shift variable is zero), the average age is minimized when $k=1$. However, this is not the case when updates arrive exogenously. We observe in Fig.~\ref{fig:buildblock}(b) that the age minimizing $k$ value can be greater than 1, and it depends on the update arrival rate, $\mu$. As $\mu$ increases, the optimal $k$ decreases and approaches 1. The reason for this is that the random waiting time with exogenous arrivals introduces a random shift to the exponential distribution of the service time. 

\section{Two-Hop Network} \label{twohop_netw}
Using the building block problem solved in the previous section, we are now ready to solve the two-hop problem ($L=2$) described in Section~\ref{model} as a preliminary step towards solving the most general case for arbitrary $L$. In the two-hop system, middle nodes cannot generate updates rather they receive them from the source node. Thus, each middle node and its $n$ children nodes in the second hop can be modeled as in Section~\ref{section:building-block}. Since the source node sends updates to the first $k_1$ of its nodes, a middle node receives a certain update packet with probability $p_1 = \frac{k_1}{n}$. If a middle node receives update $j$ and the next one it receives is update $j+M_1$, as in the building block problem, $M_1$ is a geometrically distributed random variable with parameter $p_1$. 

Since the source generates a new update at will once the current update is delivered to $k_1$ middle nodes, the interarrival time between updates that start service in the first hop is $Y_1 = X_{k_1:n}$. Let random variable $S_1$ denote the interarrival time at the receiver nodes in the first hop. In other words, let $S_1$ denote the time between updates that leave service in the first hop. Since each update cycle takes $X_{k_1:n}$ units of time, successful interarrival time at the middle nodes can be written as
\begin{align}
S_1 = \sum_{i=1}^{M_1} (Y_1)_i = \sum_{i=1}^{M_1} (X_{k_1:n})_i \label{interarrival_firsthop}
\end{align}
where the mean interarrival time is $E[S_1]=E[M_1]E[X_{k_1:n}]$. Note that we have $Z_1 = 0$ for the source node since it generates the updates as soon as the previous one is completed (transmitted to $k_1$ nodes). The receiver nodes in the first hop immediately relay the update that they have received. Therefore, the interarrival time between the successfully received updates at the receiver nodes in the first hop, $S_1$, is equal to update interarrival time at the transmitter nodes in the second hop, $R_2$.

After receiving the updates with interarrival time $R_2=S_1$, middle nodes transmit each update until it is delivered to $k_2$ of their children nodes. Similar to the first hop, when a middle node transmits an update, an end node receives the update with probability $p_2 = \frac{k_2}{n}$. Geometrically distributed $M_2$ with parameter $p_2$ denotes the number of cycles between successive updates to an end node. The interarrival time between updates that depart from a middle node and start service in the second hop is $Y_2 = \tilde{X}_{k_2:n} + Z_2$. Let random variable $S_2$ denote the interarrival time between updates that successfully arrive at the receiver nodes in the second hop. Then,
\begin{align}
S_2 = \sum_{i=1}^{M_2} (Y_2)_i = \sum_{i=1}^{M_2} (\tilde X_{k_2:n} + Z_2)_i \label{interarrival_secondhop}
\end{align}

Note that in this model a successful update reaches an end node without being preempted in each of the hops. Thus, the service time of a successful update denoted by $\bar{X}$ is the sum of link delays in each hop and corresponds to the total time spent in the system by that update. Then, the total service time of a successful update delivered to some node $i$ through middle node $j$ is $E[\bar{X}] = E[X_j | j \in  \mathcal{K}] + E[\tilde{X_i}| i \in \mathcal{K}_j]$. Here the set $ \mathcal{K}$ is the set of first $k_1$ middle nodes that receive the update and the set $ \mathcal{K}_j$ defined for each $j$ in $\mathcal{K}$ is the set of first $k_2$ end nodes that receive the update. Thus, for an update to reach an end node that end node has to be among the earliest $k_2$ children nodes of its middle node and the corresponding middle node has to be one of the earliest $k_1$ middle nodes. Now, by using (\ref{eqn:age_kmu2}), the average age of an end node for a two-hop system is given in the following theorem.
\begin{theorem} \label{thm_k1k2}
	For a two-hop system with the earliest $k_1$, $k_2$ stopping scheme, the average age at an individual end node is
	\begin{align}
	\Delta_{(k_1, k_2)} = \frac{1}{k_1} \sum_{i=1}^{k_1} E[X_{i:n}] + \frac{1}{k_2} \sum_{i=1}^{k_2} E[\tilde{X}_{i:n}] + \frac{2n-k_2}{2k_2}(E[\tilde{X}_{k_2:n}] + E[Z_2]) + \frac{Var[\tilde{X}_{k_2:n}+Z_2]}{2(E[\tilde{X}_{k_2:n}] + E[Z_2])}
	\label{eqn:thm_k1k2}
	\end{align} 
\end{theorem}

\begin{Proof}
	This theorem follows from Theorem~\ref{thm_kmu} upon observing that the second hop is the same as the building block problem. However, successful updates in the two-hop setting spend time in both hops to reach the end nodes. Noting also that successful updates do not wait in the system till they reach the end nodes, we find 
	\begin{align}
		E[\bar{X}] = \frac{1}{k_1} \sum_{i=1}^{k_1} E[X_{i:n}] + \frac{1}{k_2} \sum_{i=1}^{k_2} E[\tilde{X}_{i:n}] \label{twohop_Xbar}
	\end{align}
	Using (\ref{twohop_Xbar}) in (\ref{eqn:age_kmu2}) yields the theorem.
\end{Proof}

This theorem is valid for any distribution for $X$ and $\tilde{X}$. Random variable $Z_2$ denotes the residual interarrival time before the next update arrives to the middle node. When we no longer have exponential interarrival times, it is not easy to determine the first and second order statistics of $Z_2$. However, we can upper bound the average age of our model $\Delta_{(k_1, k_2)}$ with the average age under exponential interarrivals to the middle nodes $\Delta'_{(k_1, k_2)}$ using the following lemma.

\begin{lemma} \label{lemma_bound}
	For a two-hop system, let $\Delta'_{(k_1, k_2)}$ denote the average age of an end node under exponential interarrivals to middle nodes with mean $E[R_2]$. Then, we have $\Delta_{(k_1, k_2)} \leq \Delta'_{(k_1, k_2)}$.
\end{lemma}
The proof of Lemma~\ref{lemma_bound} follows from the DMRL (decreasing mean residual life) \cite{Marshall07} property of interarrival times and NBUE (new better than used in expectation) \cite{Marshall07} property of service times and is provided in Section~\ref{section:proof}. Note that as long as the aforementioned conditions on the interarrival and service time distributions are met, this lemma also applies to the case with $L>2$ hops. This generalization is made in Section~\ref{section:proof}.  Since $\Delta_{(k_1, k_2)} \leq \Delta'_{(k_1, k_2)}$, in order to prove that the average age in the two-hop system is bounded by a constant, all we need to show is that $\Delta'_{(k_1, k_2)}$ is upper bounded by a constant as the network grows. 

\begin{corollary} \label{corr_exp_2stage}
For a two-hop system, assuming exponential interarrivals to the middle nodes with $E[R_2]$, the average age at an end node under the earliest $k_1$,  $k_2$ stopping scheme is
	\begin{align}
	\Delta'_{(k_1, k_2)} =& \frac{1}{k_1} \sum_{i=1}^{k_1} E[X_{i:n}] + \frac{1}{k_2} \sum_{i=1}^{k_2} E[\tilde{X}_{i:n}] + \frac{2n-k_2}{2k_2}E[\tilde{X}_{k_2:n}] + \frac{2n^2-nk_2}{2k_1k_2}E[X_{k_1:n}]  \nonumber \\ &+ \frac{k_1Var[\tilde{X}_{k_2:n}]}{2(k_1E[\tilde{X}_{k_2:n}] + nE[X_{k_1:n}])} + \frac{n^2E[X_{k_1:n}]^2}{2k_1(k_1E[\tilde{X}_{k_2:n}] + nE[X_{k_1:n}])} \label{corr_exp_2stage_}
	\end{align}
\end{corollary}

\begin{Proof}
When the interarrival times to the middle nodes, $R_2$ are exponential, then $Z_2 = R_2$ due to the memoryless property of the exponential distribution. In addition, we know that $R_2 = S_1$. Therefore, $Z_2$ is  exponential with mean $E[Z_2] = E[S_1] = E[M_1]E[X_{k_1:n}]$. Then, $Var[Z_2] =E[S_1]^2 = E[M_1]^2E[X_{k_1:n}]^2$ where $M_1$ is geometrically distributed with $p_1 = \frac{k_1}{n}$. Combining these and noting that $Z_2$ and $\tilde{X}_{k_2:n}$ are independent yields the result.
\end{Proof}

\begin{corollary} \label{corr_twohop_large}
For a two-hop system, assuming $n$ is large and $n>k_1$ and $n>k_2$ and letting $k_1 = \alpha_1n$ and $k_2 = \alpha_2n$, under exponential interarrival assumption to the middle nodes with mean $E[R_2]$, and shifted exponential service times $X$ with $(\lambda,c)$ and $\tilde{X}$ with $(\tilde{\lambda}, \tilde{c})$, the average age for the earliest $k_1, k_2$ scheme can be approximated as
	\begin{align}
	%\hat{\Delta'}(\alpha_1,\alpha_2)
	\Delta'_{(k_1,k_2)}  \approx& \frac{1}{\lambda} + \frac{1}{\tilde{\lambda}} + \frac{\tilde{c}}{\alpha_2} + \frac{\tilde{c}}{2} -\frac{1}{2\tilde{\lambda}}\log(1-\alpha_2) + \frac{2-\alpha_2+2\alpha_1\alpha_2}{2\alpha_1\alpha_2}c  \nonumber \\ 
	& + \frac{\tilde{\lambda}K_1^2}{2\alpha_1\lambda[\lambda\alpha_1K_2+\tilde{\lambda}K_1]} +  \frac{3\alpha_2-2\alpha_1\alpha_2-2}{2\alpha_1\alpha_2\lambda}\log(1-\alpha_1)  \label{twohop-age}
	\end{align}	
	where $K_1 = (\lambda c-\log(1-\alpha_1))$ and $K_2 = (\tilde{\lambda}\tilde{c}-\log(1-\alpha_2))$.
\end{corollary}

With Corollary~\ref{corr_twohop_large} we have showed that $\Delta'_{(k_1, k_2)}$ derived in Corollary~\ref{corr_exp_2stage} is independent of $n$ for large $n$. Since it upper bounds our age expression $\Delta_{(k_1, k_2)}$, we conclude that age under the earliest $k_1$, $k_2$ stopping scheme for a two-hop multicast network is also independent of $n$ for large $n$ and is bounded by a constant as the number of end nodes increases.

\section{Extension to $L$ Hops}
In this section, we extend our two-hop age results in (\ref{eqn:thm_k1k2}), (\ref{corr_exp_2stage_}), and (\ref{twohop-age}) to $L$ hops. Considering an $L$-hop network, we have a single source node, $n$ first hop receiver nodes, $n^2$ second hop receiver nodes and extending in this manner, $n^L$ end ($L$ hop) nodes. The network model for $L=2$ is shown in Fig.~\ref{fig:model} and it is generalized such that each of $n$ nodes of the first hop is tied to $n$ further nodes making $n^2$ second hop nodes and similarly, each of these $n^2$ second hop nodes is further connected to $n$ nodes forming $n^3$ end nodes, and so on, for $L$ hops. 

Remember that in the two-hop model, we denote the first hop link delays as $X$, and the second hop link delays as $\tilde{X}$. In this section, in order to accommodate general $L$ hops, we change our notation so that the first hop link delay is now denoted as $X^{(1)}$, the second hop link delay is now denoted as $X^{(2)}$, the $\ell$th hop is denoted as $X^{(\ell)}$, and the last hop link delay is denoted as $X^{(L)}$. For each hop $\ell$, we utilize the earliest $k_\ell$ transmission policy such that once hop $\ell-1$ receiver nodes receive an update they begin to act as hop $\ell$ transmitter nodes and relay the update they have received to $k_\ell$ of their $n$ children nodes (see Fig.~\ref{fig:1.5}). After the packet transmission to $k_\ell$ children nodes are completed, hop $\ell$ transmitter nodes start waiting for the next update. Here, random variable $Z_\ell$ denotes this waiting time upon the completion of an update until the next one arrives. Thus, the interarrival time between two consecutive updates that depart from the transmitter and start service in hop $\ell$ is $Y_\ell = {X}_{k_\ell:n}^{(\ell)} + Z_\ell$.

Overall this $L$-hop network implements the earliest $\{k_\ell\}_{\ell=1}^L$ transmission scheme. At each hop, we have random variable $M_\ell$ which is geometrically distributed with parameter $p_\ell = \frac{k_\ell}{n}$ that represents the number of update cycles between two successive updates that arrive at a receiver in hop $\ell$. The interarrival time between two consecutive updates that leave service in hop $\ell$ without being preempted can be written as 
\begin{align}
S_\ell = \sum_{i=1}^{M_\ell} (Y_\ell)_i = \sum_{i=1}^{M_\ell} ({X}_{k_\ell:n}^{(\ell)} + Z_\ell) \label{interarrival_Lhop}
\end{align}
A receiver node in hop $\ell$ immediately transmits an update it receives to its children nodes in hop $\ell+1$. Therefore, the interarrival time between two consecutive successful updates that leave service in hop $\ell$, $S_\ell$, is equal to the interarrival time between two consecutive updates that arrive in hop $\ell+1$, $R_{\ell+1}$. We have $R_{\ell+1} = S_{\ell}$. 

The last hop in this $L$-hop network can be seen as an application of our building block problem. In the building block problem, there is a source node and an end node. When this is applied to the last hop of the $L$-hop network, the source node is an arbitrary receiver node in hop $L-1$, and the end node is an arbitrary receiver node in hop $L$. Interarrival time of updates that arrive exogenously at the source node in the building block problem, $R$, is now $R = S_{L-1}$, and the interarrival time between successful updates that arrive at the end node in the building block problem, $S$, is now $S = S_L$. In addition, we know that $S_L = \sum_{i=1}^{M_L} (Y_L)_i$. Finally, the service time in the building block problem, $\bar{X}$, is now $\bar{X}= \sum_{\ell=1}^L \bar{X}_\ell$, where $\bar{X}_\ell$ is the service time at each hop. We know that for an update packet to reach the end nodes, it has to be among the earliest $k_1$, \dots, $k_L$ nodes in all hops. Thus, $\sum_{\ell=1}^L \bar{X}_\ell$ term captures the expected time spent in the system for a successful update without being preemopted until it reaches one of the end nodes. The average age of an end node for an $L$-hop system is given in the following theorem.

\begin{theorem} \label{thm_kL}
For the general $L$-hop network with the earliest $\{k_\ell\}_{\ell=1}^{L}$ stopping scheme, the average age at an individual end node at hop $L$ is
\begin{align}
\Delta_{\{k_\ell\}_{\ell=1}^{L}} =& \sum_{\ell=1}^L \frac{1}{k_\ell} \sum_{i=1}^{k_\ell} E[X_{i:n}^{(\ell)}] + \frac{2n-k_L}{2k_L}(E[{X}_{k_L:n}^{(L)}] + E[Z_L]) + \frac{Var[{X}_{k_L:n}^{(L)}+Z_L]}{2(E[{X}_{k_L:n}^{(L)}] + E[Z_L])} \label{threehop_corr}
\end{align}
\end{theorem}
\begin{Proof}
Since the last hop of the general $L$-hop network can be seen as a building block, we can apply (\ref{eqn:age_kmu2}) to the setting here by inserting $Y = Y_L$ and $\bar{X}= \sum_{\ell=1}^L \bar{X}_\ell$. Then, (\ref{threehop_corr}) follows after similar calculations as in Theorems \ref{thm_kmu} and \ref{thm_k1k2}.
\end{Proof}

Similar to Corollary~\ref{corr_exp_2stage}, the age of the last hop can be upper bounded by assuming that the interarrival times at each hop are exponentially distributed. 
\begin{corollary} \label{the:Lstage}
For the general $L$-hop network, assuming exponential interarrivals to each hop with mean $E[R_\ell]$, average age at an end node under $\{k_\ell\}_{\ell=1}^{L}$ stopping scheme is given by
	\begin{align}
	\Delta'_{\{k_\ell\}_{\ell=1}^{L}} =& \sum_{\ell=1}^L \frac{1}{k_\ell} \sum_{i=1}^{k_\ell} E[X_{i:n}^{(\ell)}] +
	\frac{2n-k_L}{2k_L} \sum_{\ell=1}^L E[X_{k_\ell:n}^{(\ell)}]\prod_{i=\ell}^{L-1}E[M_i]  \nonumber \\
	&+ \frac{Var[{X}_{k_L:n}^{(L)}] + \left(\sum_{\ell=1}^{L-1} E[X_{k_\ell:n}^{(\ell)}]\prod_{i=\ell}^{L-1}E[M_i]\right)^2}{2(\sum_{\ell=1}^L E[X_{k_\ell:n}^{(\ell)}]\prod_{i=\ell}^{L-1}E[M_i])} \label{eqn:TheoremL}
	\end{align}
\end{corollary}
\begin{Proof}
Using Lemma~\ref{lemma_bound}, (\ref{threehop_corr}) can be upper bounded with exponential interarrivals to the nodes at hop $L-1$, where mean interarrival time is $E[R_L] = E[S_{L-1}]$. When the interarrivals are exponentially distributed, we have $Z_L = S_{L-1}$ and therefore, $E[Z_L] = E[S_{L-1}]$. In addition, with exponential interarrivals, $Z_L$ is independent of ${X}_{k_L:n}$. Then, we have $Var[{X}_{k_L:n}^{(L)}+Z_L] = Var[{X}_{k_L:n}^{(L)}]+Var[Z_L]$, where $Var[Z_L] = E[S_{L-1}]^2$. Now, the upper bound for (\ref{threehop_corr}) can be written as
\begin{align}
\Delta'_{\{k_\ell\}_{\ell=1}^{L}} =& \sum_{\ell=1}^L \frac{1}{k_\ell} \sum_{i=1}^{k_\ell} E[X_{i:n}^{(\ell)}] + \frac{2n-k_L}{2k_L}(E[{X}_{k_L:n}^{(L)}] + E[S_{L-1}]) + \frac{Var[{X}_{k_L:n}^{(L)}]+E[S_{L-1}]^2}{2(E[{X}_{k_L:n}^{(L)}] + E[S_{L-1}])} \label{eqn:threehop_upper}.
\end{align}

Now, let us calculate $E[S_{L-1}]$. We can write $S_{L-1}$ as
\begin{align}
S_{L-1} =  \sum_{i=1}^{M_{L-1}} (Y_{L-1})_i = \sum_{i=1}^{M_{L-1}} ({X}_{k_{L-1}:n}^{(L-1)}+ Z_{L-1})_i \label{interarrival_L-1hop}
\end{align}
where $Y_{L-1}$ is the interarrival time between updates that start service in hop $L-1$. Then,
\begin{align}
E[Z_L] = E[S_{L-1}] = E[M_{L-1}]E[{X}_{k_{L-1}:n}^{(L-1)}]+E[M_{L-1}]E[Z_{L-1}]. \label{EZ_3}
\end{align}
Similarly, $E[Z_{L-1}]$ can be written in terms of the variables in hop $L-2$. We continue with this recursive calculation until the second hop, where we have $E[Z_2] = E[M_1]E[X_{k_1:n}^{(1)}]$. As a result, we have
\begin{align}
E[S_{L-1}] =& E[M_{L-1}]E[{X}_{k_{L-1}:n}^{(L-1)}] + \cdots + E[M_{L-1}]E[M_{L-2}]\cdots E[M_2]E[M_1]E[{X}_{k_{1}:n}^{(1)}] \\
= & \sum_{\ell=1}^{L-1} E[X_{k_\ell:n}^{(\ell)}]\prod_{i=\ell}^{L-1}E[M_i]  \label{eqn:ZL}
\end{align} 
Adding $E[{X}_{k_{L-1}:n}^{L-1}]$ to (\ref{eqn:ZL}), we have
\begin{align}
E[{X}_{k_{L-1}:n}^{L-1}] + E[S_{L-1}] = \sum_{\ell=1}^{L} E[X_{k_\ell:n}^{(\ell)}]\prod_{i=\ell}^{L-1}E[M_i]. \label{eqn:ZL+}
\end{align} 
Finally, by inserting (\ref{eqn:ZL}) and (\ref{eqn:ZL+}) into (\ref{eqn:threehop_upper}), we obtain (\ref{eqn:TheoremL}).
\end{Proof}

Corollary~\ref{the:Lstage} introduces an upper bound for average age of an update that is generated at the source node and delivered to any one of the end nodes in hop $L$. Note that the total number of nodes in the $L$-hop network is $\sum_{\ell=0}^L n^l$. The result of Corollary~\ref{the:Lstage} holds for any $n$. Next, we characterize the effect of increasing $n$ on the scaling of average age.

\begin{corollary}
	Assume $n$ is large and $n>k_\ell$ and let $k_\ell = \alpha_\ell n$ for $\ell=1,\dots, L$. Under exponential interarrival assumption to each hop with mean $E[R_\ell]$, and shifted exponential service times $X^{(\ell)}$ with $(\lambda_\ell, c_\ell)$, the average age for the earliest $\{k_\ell\}_{\ell=1}^{L}$ scheme can be approximated as
	\begin{align}
	\Delta'_{\{k_\ell\}_{\ell=1}^{L}}  \approx& \sum_{\ell=1}^L \left( c_\ell + \frac{1}{\lambda_\ell} + \frac{1-\alpha_\ell}{\alpha_\ell \lambda_\ell} \log (1-\alpha_\ell) \right) + \frac{2-\alpha_L}{2\alpha_L}  \sum_{\ell=1}^L \left[   \left( {c_\ell} - \frac{1}{{\lambda_\ell}} \log (1-\alpha_\ell) \right)  \prod_{i=\ell}^{L-1} \frac{1}{\alpha_i} \right] \nonumber \\
	&+ \frac{\left(  \sum_{\ell=1}^{L-1} \left[   \left( {c_\ell} - \frac{1}{{\lambda_\ell}} \log (1-\alpha_\ell) \right)  \prod_{i=\ell}^{L-1} \frac{1}{\alpha_i} \right]  \right)^2}{2\sum_{\ell=1}^L \left[   \left( {c_\ell} - \frac{1}{{\lambda_\ell}} \log (1-\alpha_\ell) \right)  \prod_{i=\ell}^{L-1} \frac{1}{\alpha_i} \right]}
	\label{threehop-age}
	\end{align}	
\end{corollary}
The proof of this corrollary is similar to that of Corrollaries~\ref{corr2} and \ref{corr_twohop_large}. We use the first and second moments of order statistics of shifted exponential random variables and make necessary approximations for large $n$.

In conclusion, under the assumption that middle nodes receive update packets with exponential interarrivals with means equal to $E[R_\ell]$, the average age attained at the end nodes depends on $n$ only through ratios $\alpha_\ell$. Thus, this result together with Lemma~\ref{lemma_bound} imply that the average age of an end node in an $L$-hop multicast network implementing the earliest $\{k_\ell\}$ transmission scheme is bounded by a constant as $n$ increases.

\section{Numerical Results} \label{num_results}

\begin{table}[t]
	\begin{center}
		\begin{tabular}{ | l | l | l | l | l |}
			\hline
			number of hops $L$ & $\alpha_1^*$ & $\alpha_2^*$ & $\alpha_3^*$ & $\alpha_4^*$ \\ \hline
			$L=1$ & $0.732$ & - & - & -  \\ \hline
			$L=2$  & $0.615$ & $0.921$ & - & - \\ \hline
			$L=3$  & $0.626$ & $0.832$ & $0.965$ & -  \\\hline
			$L=4$  & $0.635$ & $0.837$ & $0.901$ & $0.935$  \\\hline
		\end{tabular}
	\end{center}
	\caption{Optimal $\alpha_i$ values for $L$-hop network when all link delays are shifted exponentials with $(1,1)$.}
	\label{table:opt_alphas}
\end{table}

In this section, we provide simple numerical results. In order to optimize the age, we need to select appropriate $k_i$ values, i.e., optimum ratios $\alpha_i^*$, at each hop. From \cite{Zhong17a}, we know that in the single-hop multicast network $\alpha^*$ is $0.732$ when link delays are shifted exponential with parameters $(\lambda,c) = (1,1)$. For the two-hop network with $(\lambda,c) = (\tilde{\lambda}, \tilde{c}) = (1,1)$, we obtain $\alpha_1^* = 0.615$ and $\alpha_2^* = 0.921$. This shows that when all link delays are statistically identical, to achieve a good age performance, we need to be more aggressive in the second hop than the first hop (see \cite{Buyukates18}). 

 When we add a third hop, we see that the optimal threshold results follow similar trends. Let us denote the link delay parameters of the third hop as $(\tilde{\tilde{\lambda}}, \tilde{\tilde{c}})$ for tractability. When all link delays are statistically identical, i.e., shifted exponentials with parameters (1,1), we have $\alpha_1^* = 0.626 $, $\alpha_2^* = 0.832$ and $\alpha_3^* = 0.965$. In a similar fashion, when we add a fourth hop, we have  $\alpha_1^* = 0.635 $, $\alpha_2^* = 0.837$, $\alpha_3^* = 0.901$ and $\alpha_4^* = 0.935$. Thus, the observation we have made for two-hop multicast networks, i.e., that we need to be more agressive in the further hops to achieve a lower average age holds true for four-hop multicast networks as well. These results are summarized in Table~\ref{table:opt_alphas}.

\begin{figure}
	\centering
	\subfloat[\label{fig:d3}]{\includegraphics[width=.5\columnwidth]{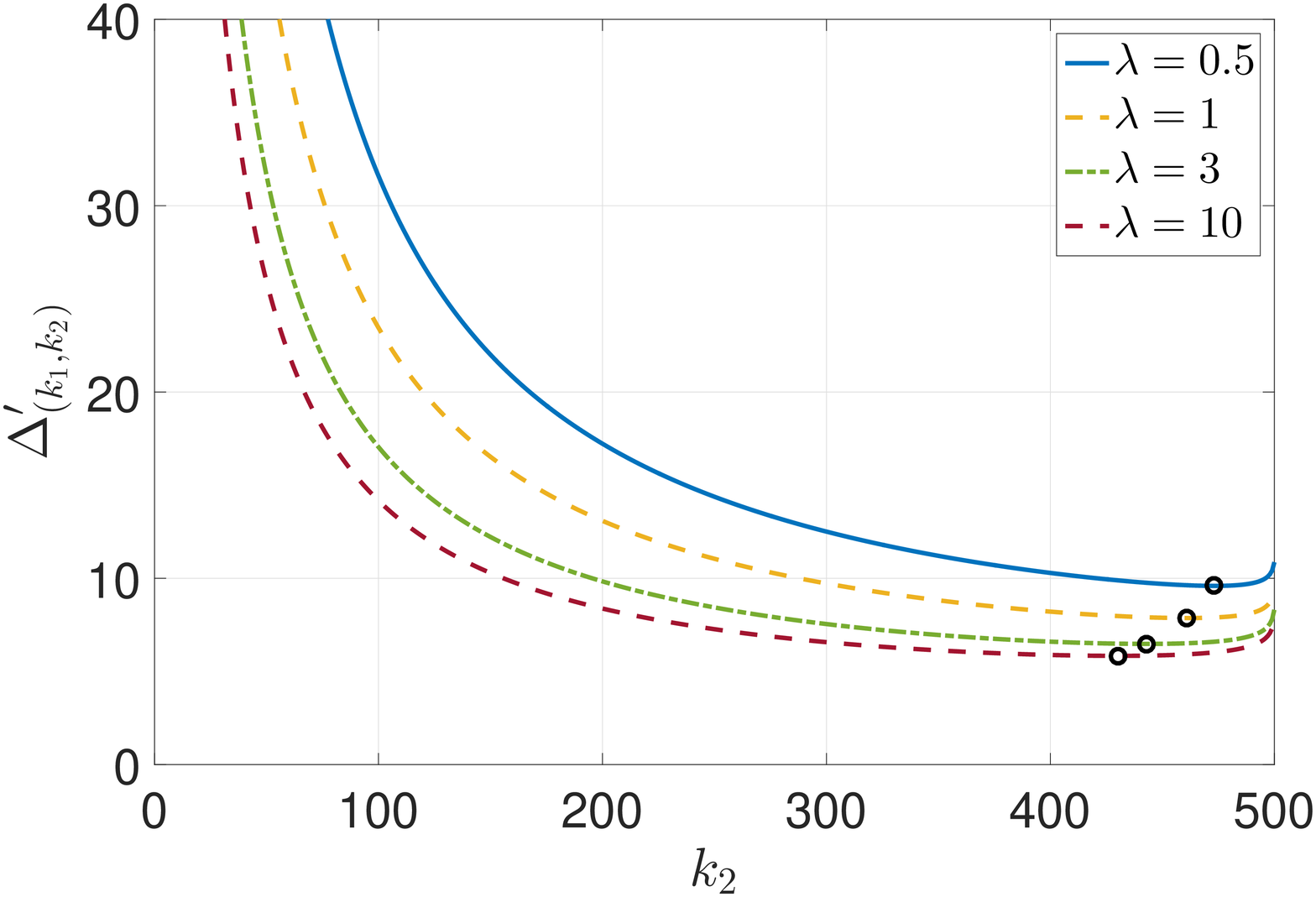} }%
	\subfloat[\label{fig:d4}]{\includegraphics[width=.5\columnwidth]{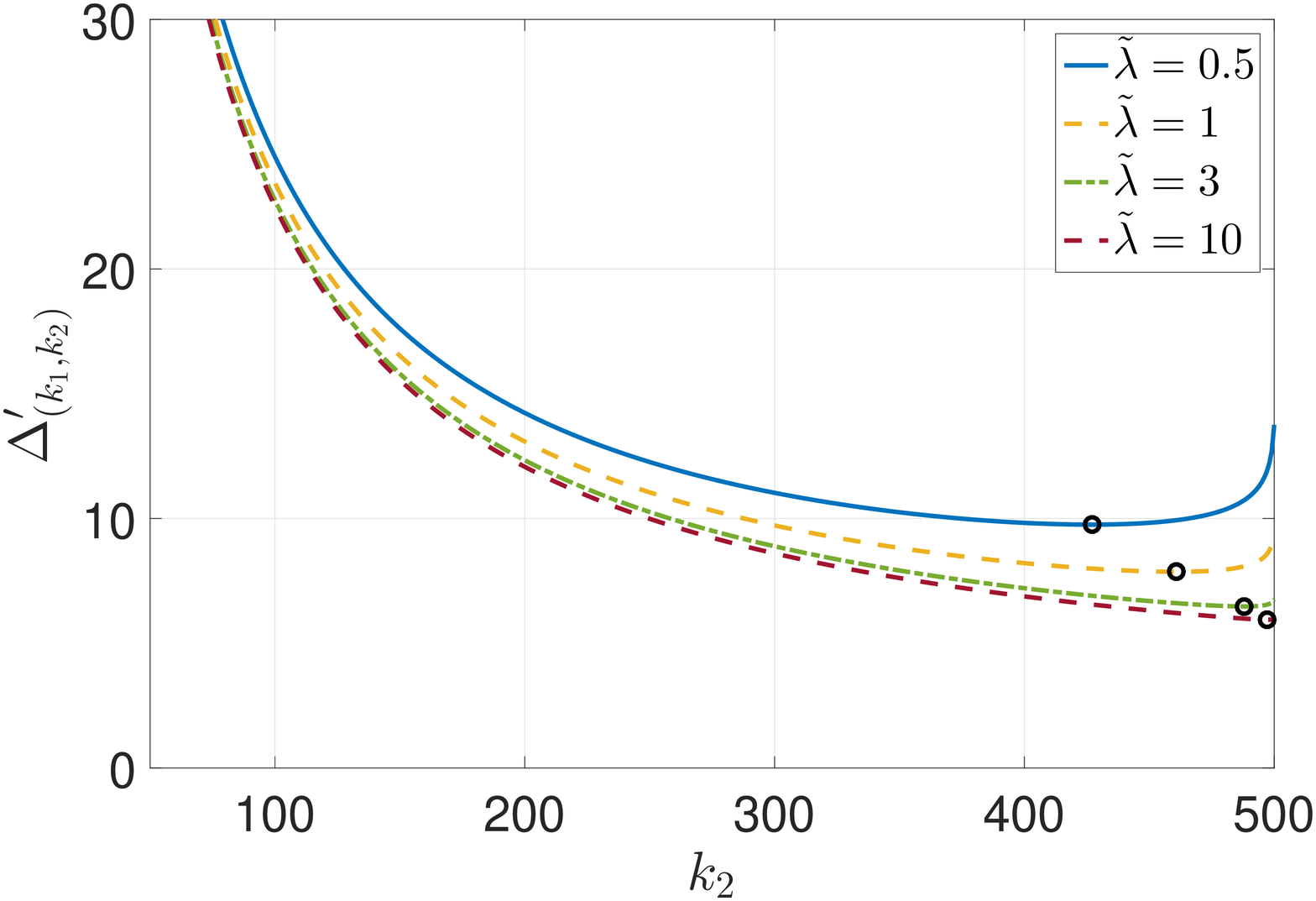} }
	\caption{$\Delta'_{(k_1,k_2)}$ as a function of $k_2$ for $n = 500$. $\circ$ marks the minimized average age $\Delta'_{(k_1,k_2)}$: (a) $c=1$ and $(\tilde{\lambda}, \tilde{c}) = (1,1)$ for varying $\lambda$, (b) $\tilde{c}=1$ and $(\lambda, c) = (1,1)$ for varying $\tilde{\lambda}$.}
	\label{fig:secondstage_k2}
\end{figure}

\begin{figure}
	\centering
	\subfloat[\label{fig:d7}]{   \includegraphics[width=.5\columnwidth]{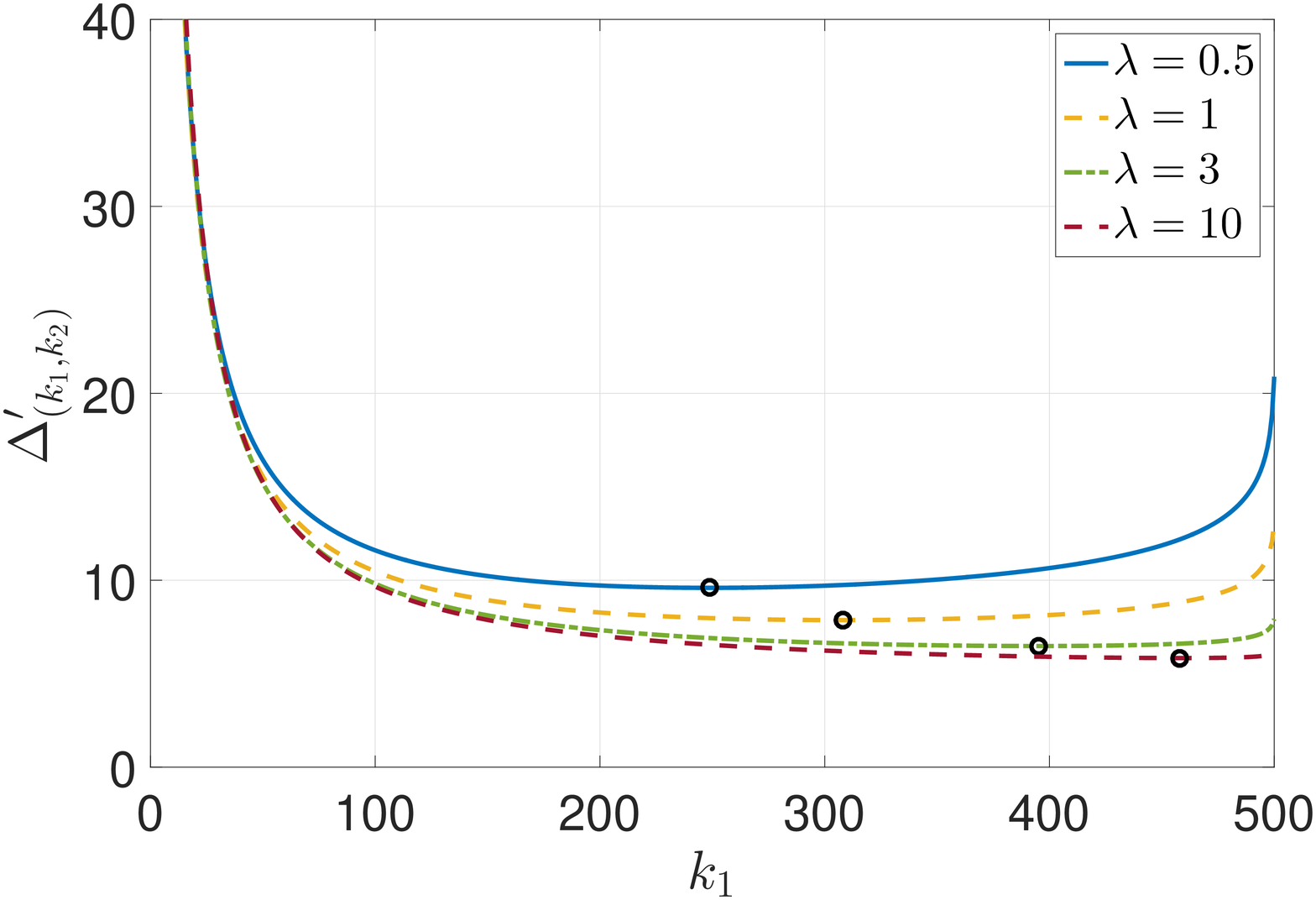}     }%
	\subfloat[\label{fig:d8}]{  \includegraphics[width=.5\columnwidth]{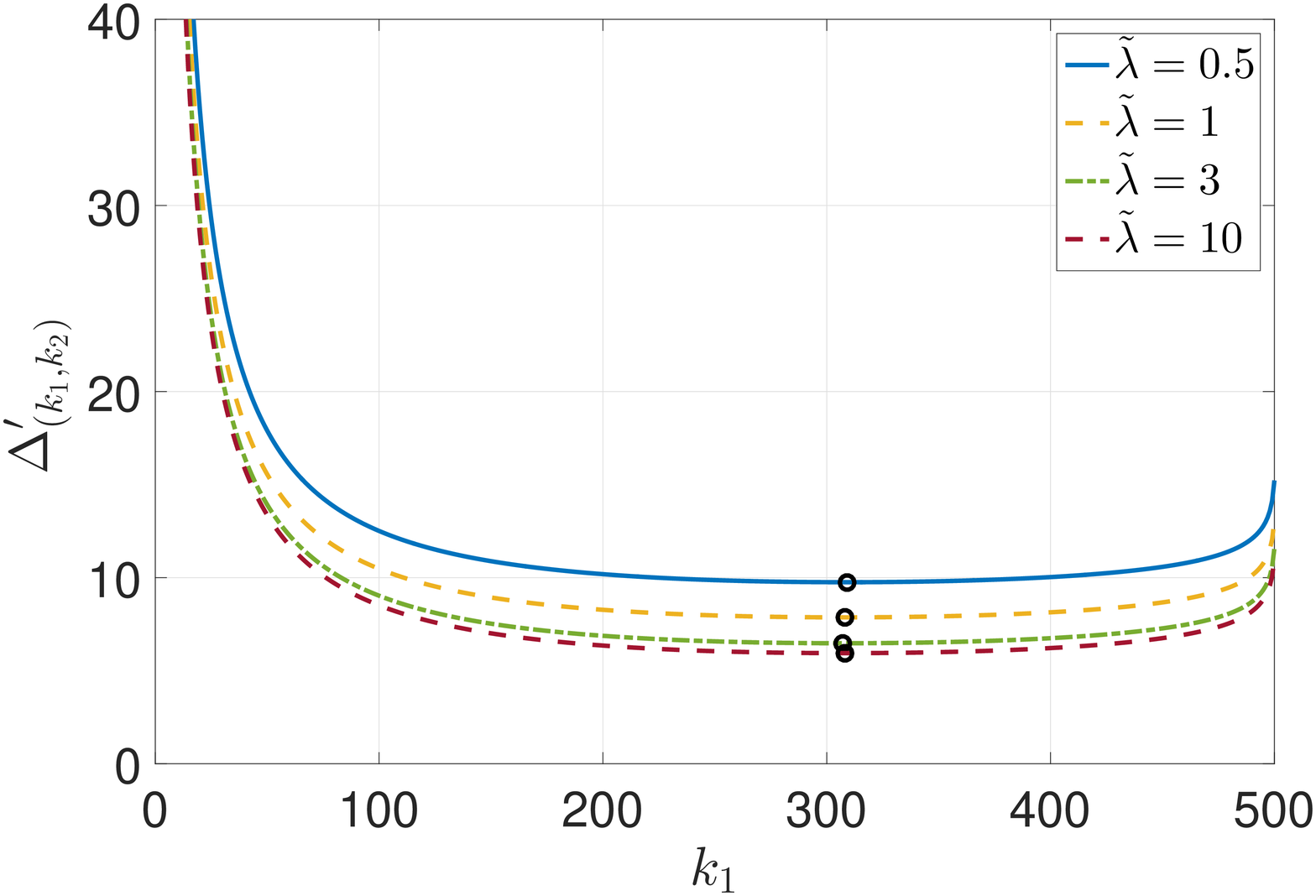}  }
	\caption{$\Delta'_{(k_1,k_2)}$ as a function of $k_1$ for $n = 500$. $\circ$ marks the minimized average age $\Delta'_{(k_1,k_2)}$: (a) $c=1$ and $(\tilde{\lambda}, \tilde{c}) = (1,1)$ for varying $\lambda$, (b) $\tilde{c}=1$ and $(\lambda, c) = (1,1)$ for varying $\tilde{\lambda}$.}
	\label{fig:secondstage_2k2}
\end{figure}

Returning to the two-hop network, we observe that $\alpha_2$ is responsive to the changes in the parameters of the first hop. This is intuitive because as $k_1$ varies, the mean interarrival time for the second hop changes. In Figs.~\ref{fig:secondstage_k2}(a) and \ref{fig:secondstage_k2}(b), we plot the age as a function of $k_2$ (equivalently $\alpha_2$) for a fixed set of second hop parameters and for a fixed set of first hop parameters, respectively. As shown in Fig.~\ref{fig:secondstage_k2}(a), for the same $(\tilde{\lambda}, \tilde{c})$ pair, when the mean interarrival time gets lower by increasing $\lambda$, $\alpha_2^*$ gets lower as well. Knowing that the next update arrival is not far away, middle nodes tend to wait for the next one instead of sending the current packet to more and more end users when the arrivals are frequent. This is exactly what we have observed in the building block problem with exogenous update arrivals. 

\begin{figure}
	\centering
	\subfloat[\label{fig:d9}]{    \includegraphics[width=.5\columnwidth]{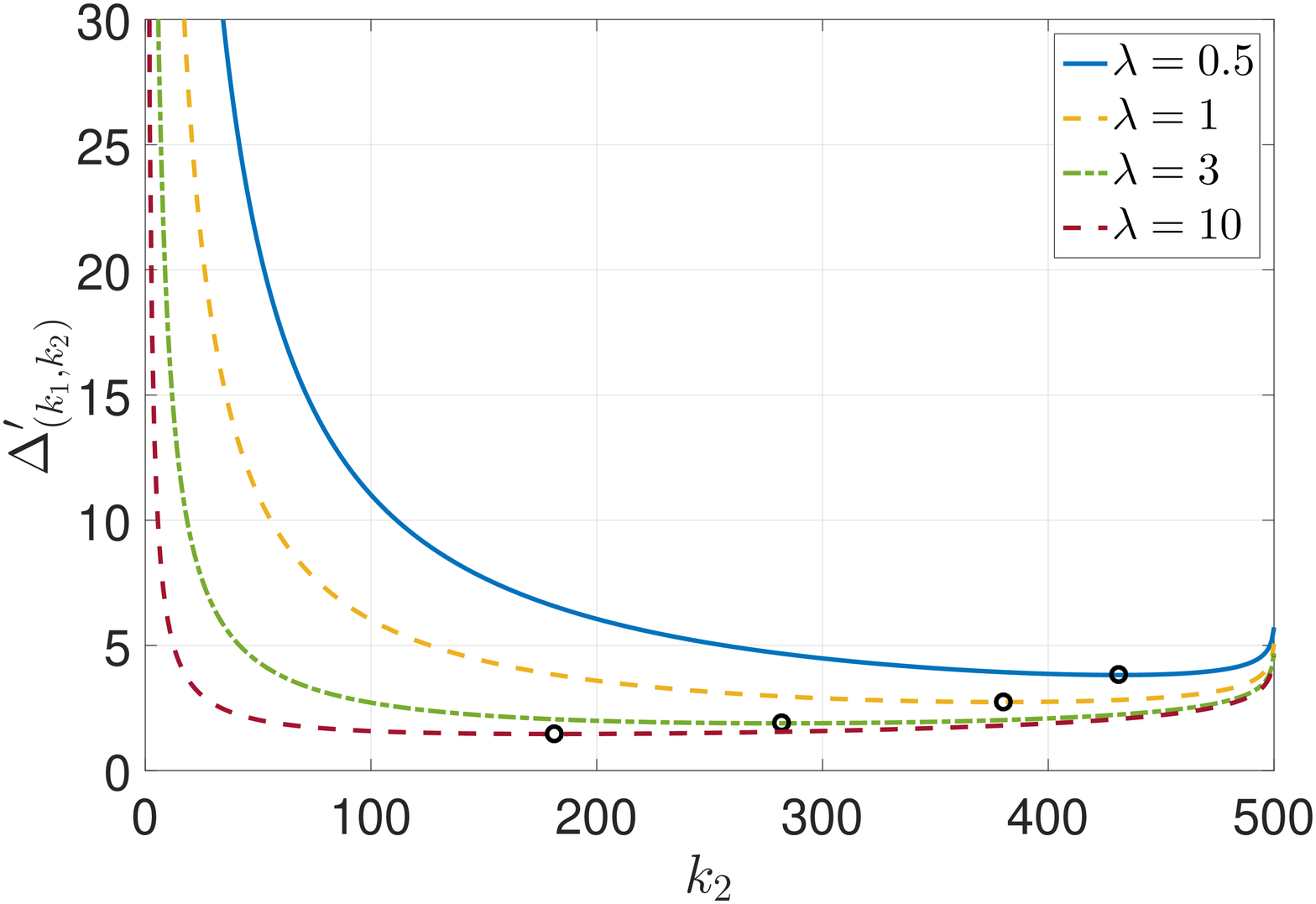}    }%
	\subfloat[\label{fig:d10}]{  \includegraphics[width=.5\columnwidth]{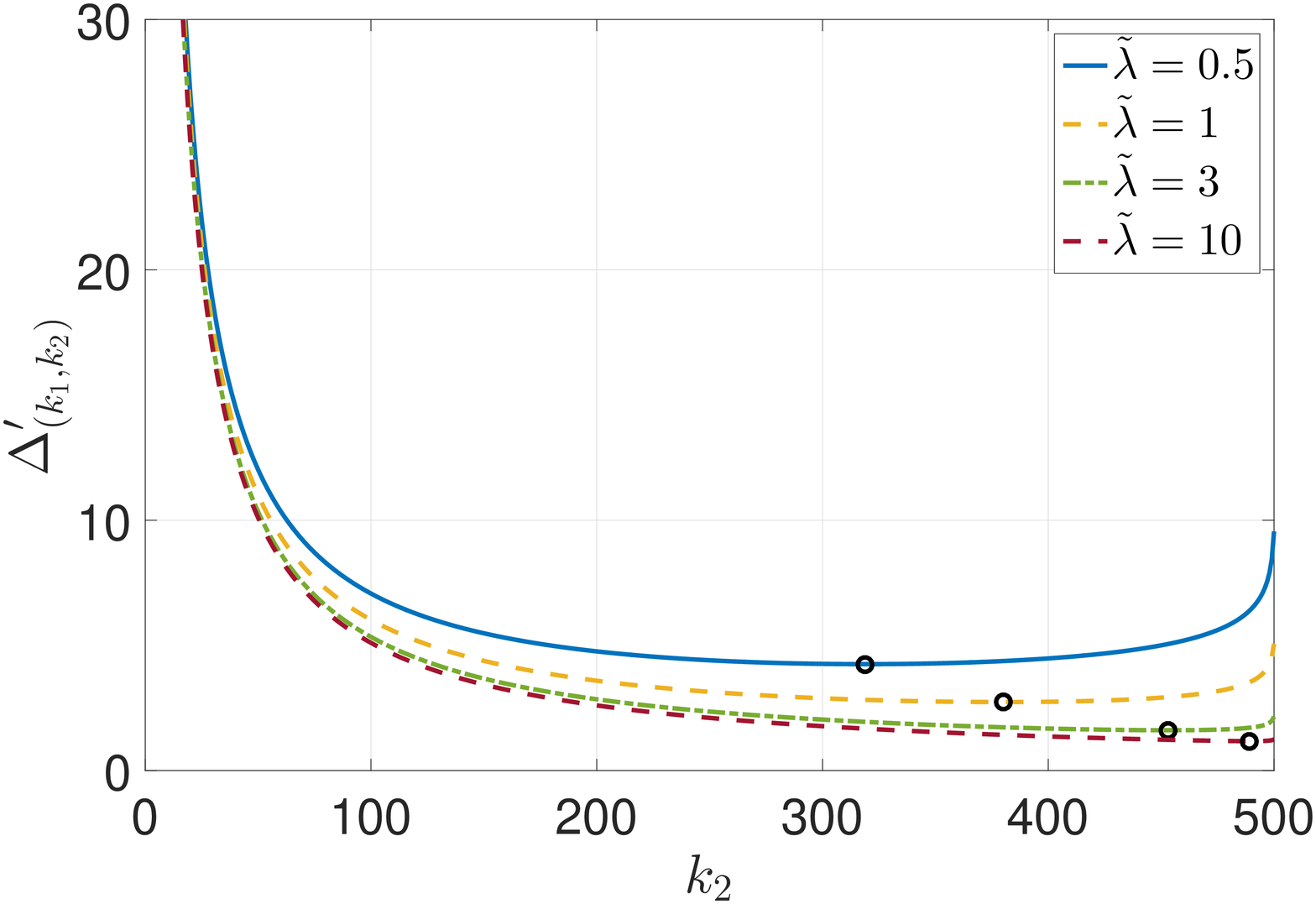}  }
	\caption{$\Delta'_{(k_1,k_2)}$ as a function of $k_2$ for $n = 500$ when link delays are exponential. $\circ$ marks the minimized average age $\Delta'_{(k_1,k_2)}$: (a) $c=0$ and $(\tilde{\lambda}, \tilde{c}) = (1,0)$ for varying $\lambda$, (b) $\tilde{c}=0$ and $(\lambda, c) = (1,0)$ for varying $\tilde{\lambda}$.}
	\label{fig:secondstage_k2_exp}
\end{figure}

\begin{figure}
	\centering
	\subfloat[\label{fig:e1}]{   \includegraphics[width=.5\columnwidth]{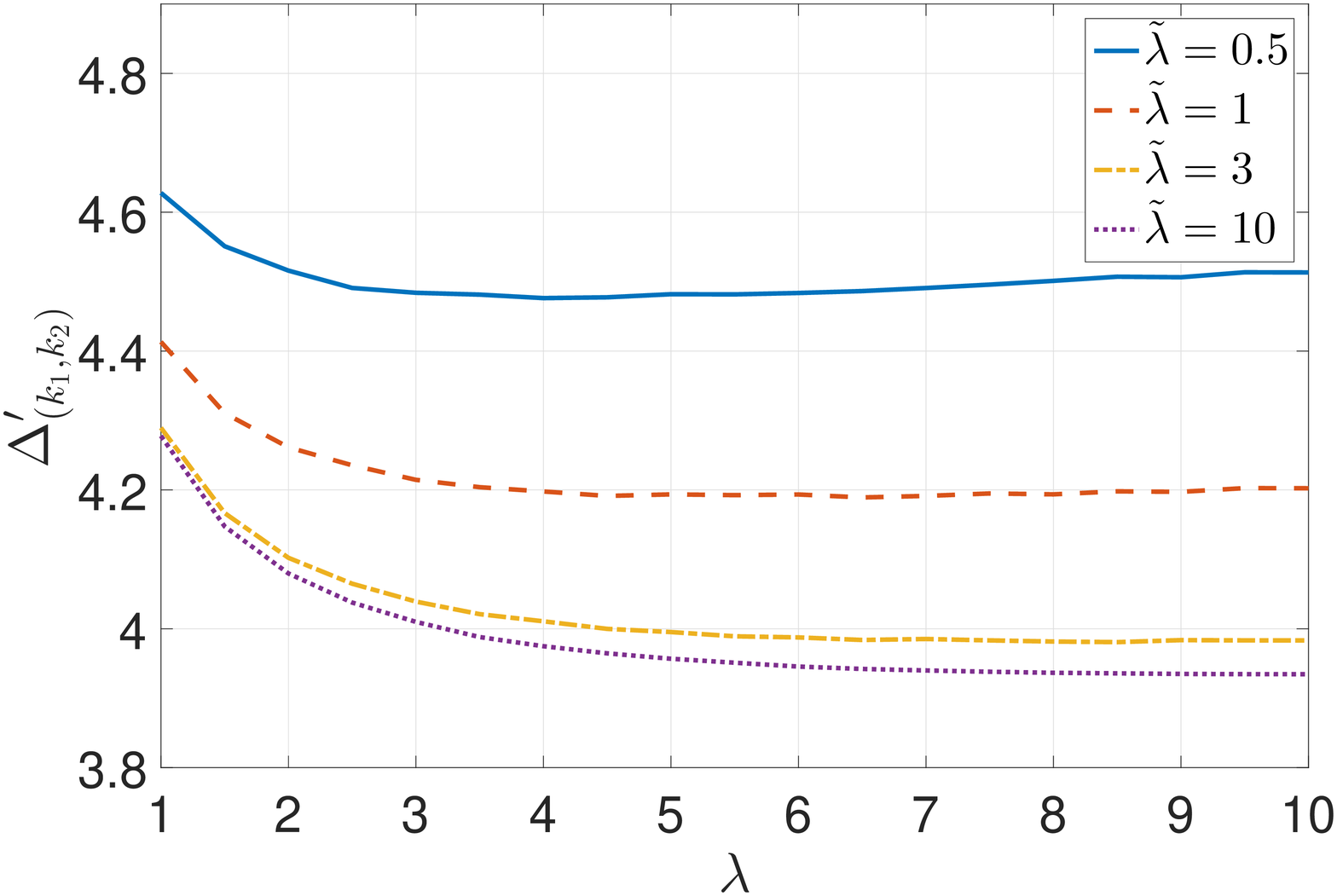}    }%
	\subfloat[\label{fig:e2}]{  \includegraphics[width=.5\columnwidth]{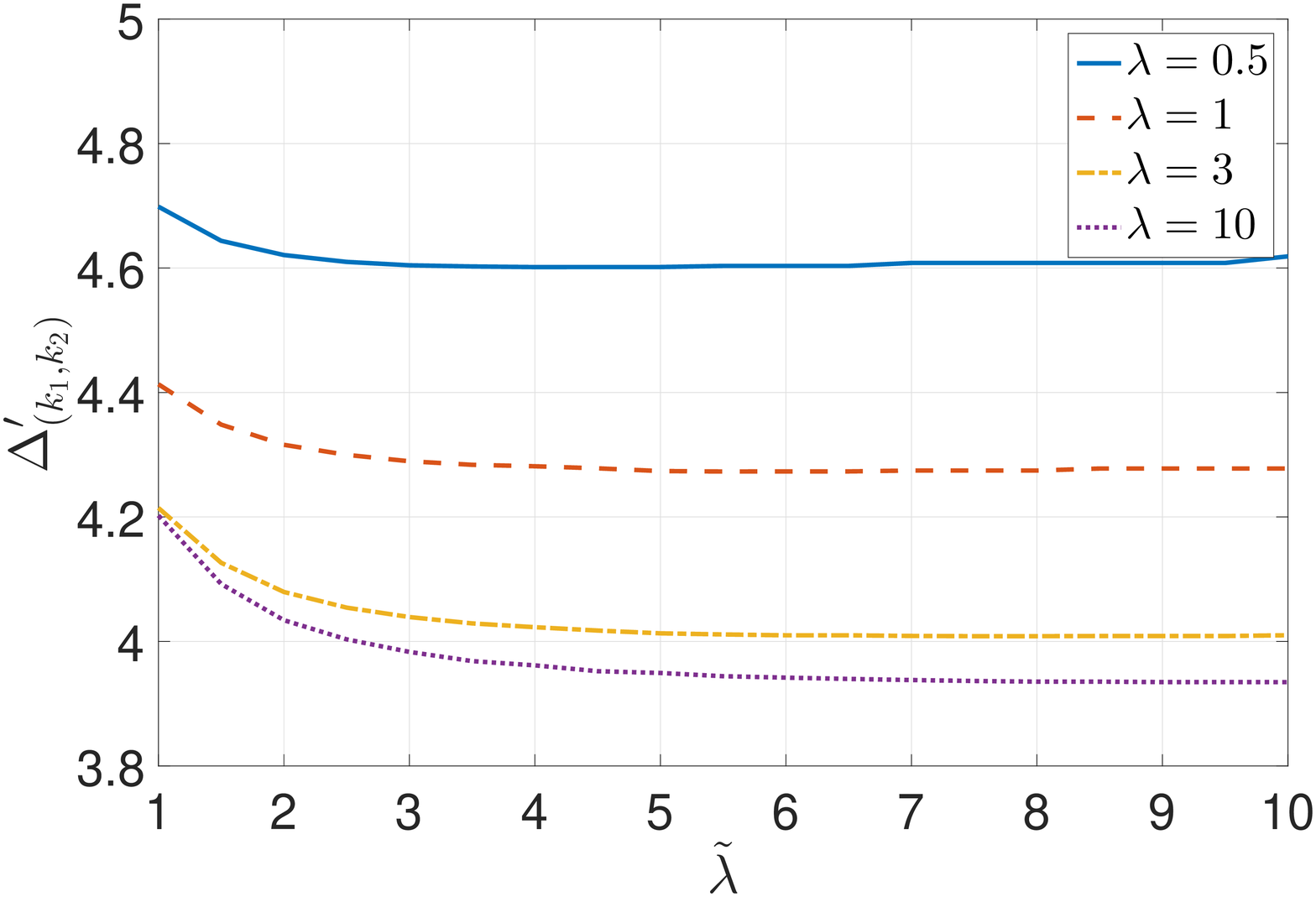}  }
	\caption{$\Delta'_{(k_1,k_2)}$ as a function of link delay parameters $\lambda$ and $\tilde{\lambda}$ for $n = 500$: (a) $\Delta'_{(k_1,k_2)}$ as a function of $\lambda$, (b) $\Delta'_{(k_1,k_2)}$ as a function of $\tilde{\lambda}$.}
	\label{fig:secondstage_lambda}
\end{figure}

In Fig.~\ref{fig:secondstage_k2}(b) we observe that as the service rate of the second hop, $\tilde{\lambda}$, increases for a given $\lambda$, $\alpha_2^*$ also increases. This is intuitive because when $\lambda$ is fixed the interarrival time to the middle nodes stays constant as opposed to their increasing service rate. Thus, they relay their current packet to more and more end nodes. 

In Figs.~\ref{fig:secondstage_2k2}(a) and \ref{fig:secondstage_2k2}(b), we plot the age as a function of $k_1$ (equivalently $\alpha_1$) when the second hop parameters are fixed and when the first hop parameters are fixed, respectively. In Fig.~\ref{fig:secondstage_2k2}(a), $\alpha_1^*$ shows a similar trend as in $\alpha_2^*$ in Fig.~\ref{fig:secondstage_k2}(b) but $\alpha_1^*$ experiences bigger changes as $\lambda$ increases. Fig.~\ref{fig:secondstage_2k2}(b) shows the response of $\alpha_1^*$ to changes in $\tilde{\lambda}$ when $\lambda$ is fixed. Here, we observe rather minor reaction from $\alpha_1^*$ when the service performance of the second stage varies. 

In Fig. \ref{fig:secondstage_k2_exp} we repeat the numerical analysis in Fig. \ref{fig:secondstage_k2} when the link delays are exponential, i.e., shift parameters $c = 0$ and $\tilde{c} = 0$. We observe similar trends for $k_2^*$ as in Fig. \ref{fig:secondstage_k2}. The only difference is now that the link delays are all exponential, we get $k_1^* = 1$ for all cases. This is also observed in \cite{Zhong17a} and in the building block problem when $\mu$ tends to $\infty$ and is because of the memoryless property of the exponential distribution.  

%When we add the third hop, we see in Fig.~\ref{fig:thirdstage_k3} that as $\lambda$ increases, meaning that as the average interarrival to the second hop decreases $\alpha_3^*$ also decreases. This is exactly what we observed for the two-hop network in Fig.~\ref{fig:secondstage_k2} and for building block problem in Fig.~\ref{fig:buildblock}.

\begin{figure}
	\centering
	\subfloat[\label{fig:d5}]{\includegraphics[width=.5\columnwidth]{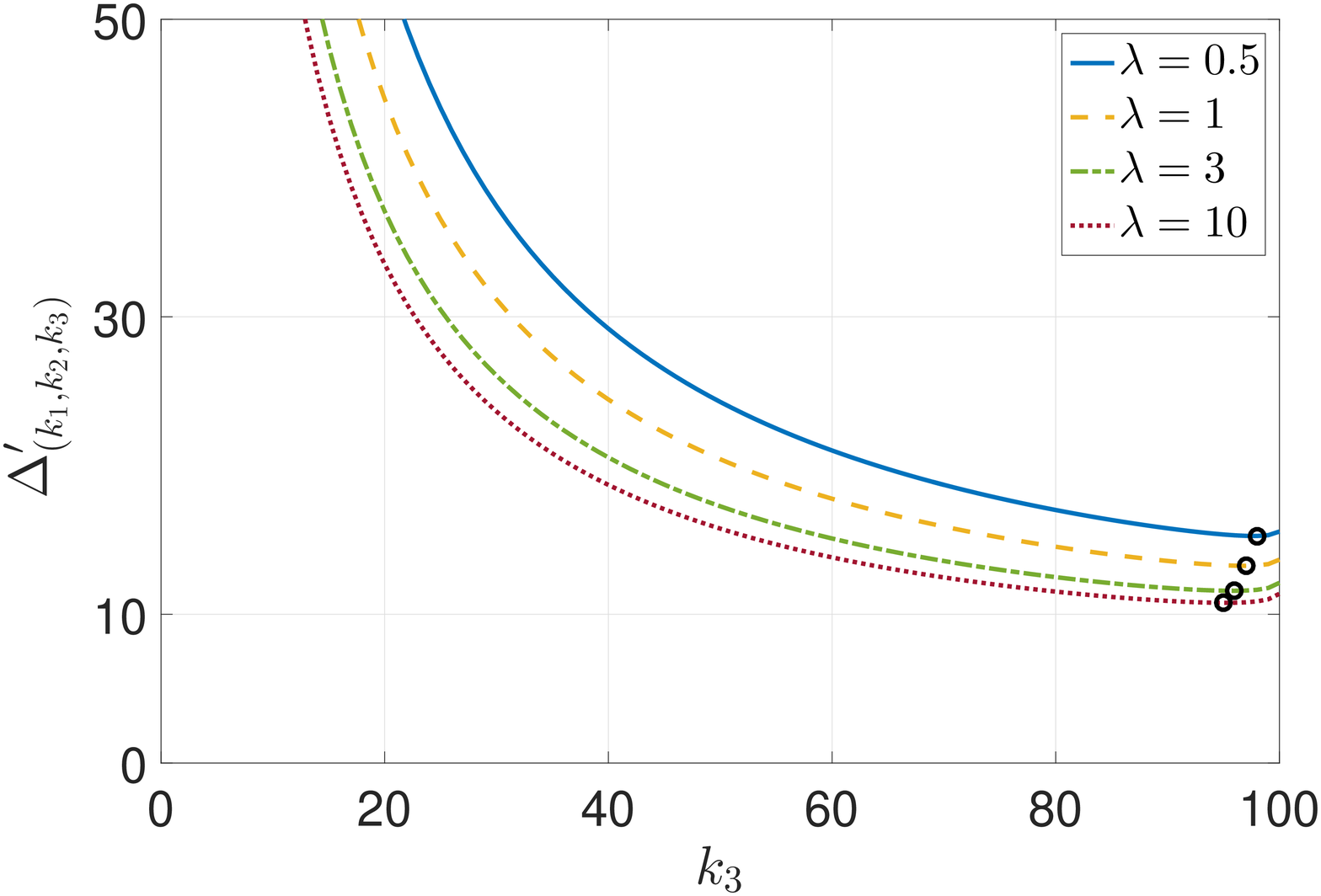}}%
	\subfloat[\label{fig:d6}]{\includegraphics[width=.5\columnwidth]{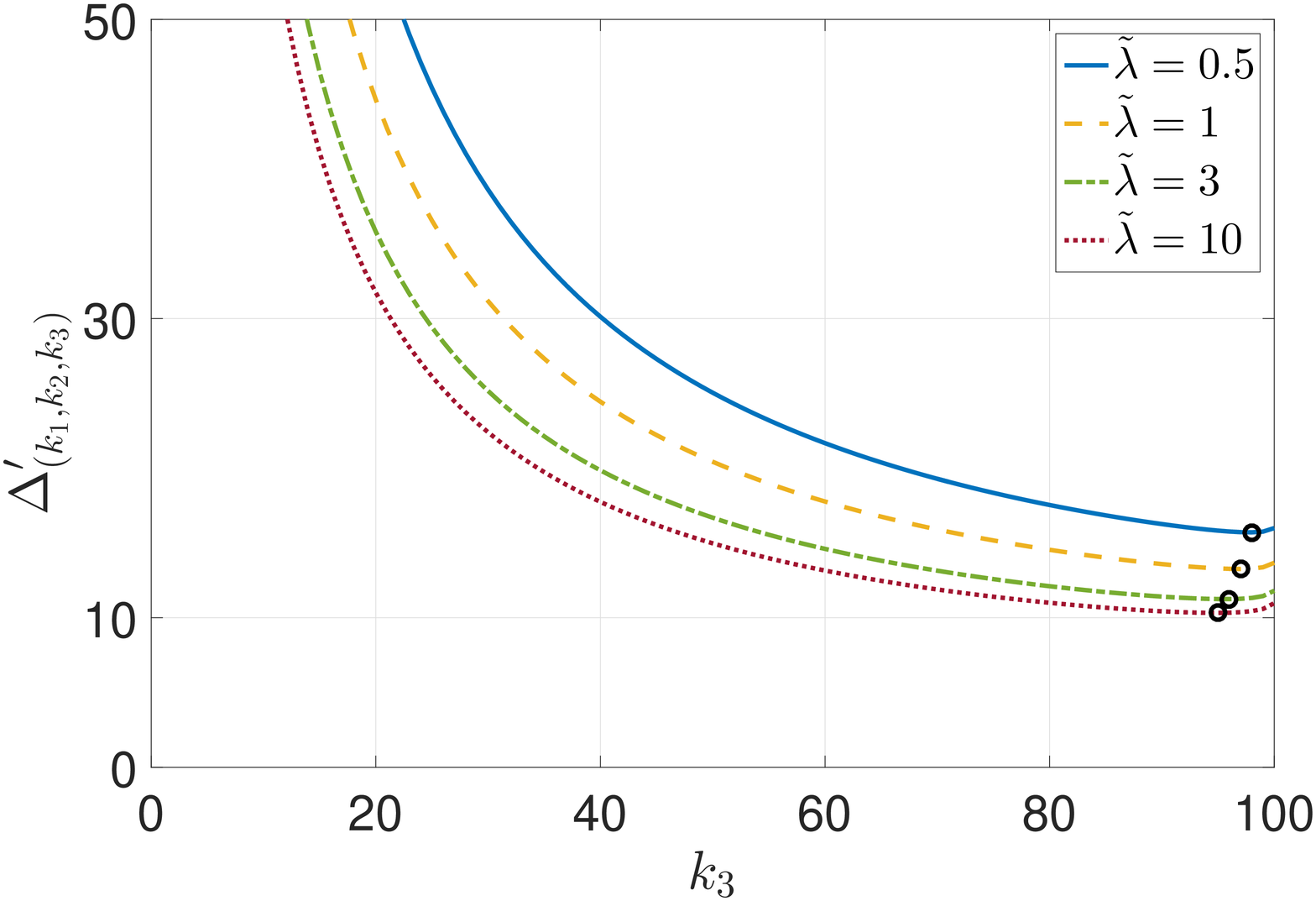} }
	\caption{$\Delta'_{(k_1,k_2,k_3)}$ as a function of $k_3$ for $n = 100$. $\circ$ marks the minimized average age $\Delta'_{(k_1,k_2,k_3)}$: (a) $c=1$ and $(\tilde{\lambda}, \tilde{c}) = (\tilde{\tilde{\lambda}}, \tilde{\tilde{c}}) = (1,1)$ for varying $\lambda$, (b) $\tilde{c}=1$ and $(\lambda, c) =  (\tilde{\tilde{\lambda}}, \tilde{\tilde{c}}) = (1,1)$ for varying $\tilde{\lambda}$.}
	\label{fig:thirdstage_k3}
\end{figure}

In Fig. \ref{fig:secondstage_lambda} we analyze the effects of the link delay parameters $\lambda$ and $\tilde{\lambda}$ on $\Delta'_{(k_1,k_2)}$. In both cases we see that larger service rates (large $\lambda$ and $\tilde{\lambda}$), i.e., lower link delays, lead to smaller average age at the end nodes. 

In Fig.~\ref{fig:thirdstage_k3} we plot the three-hop version of Fig.~\ref{fig:secondstage_k2}. As shown in Fig.~\ref{fig:thirdstage_k3}, when the mean interarrival time decreases at the third hop through a service rate increase in either one of the first two hops $\alpha_3^*$ gets smaller. One other observation from Fig.~\ref{fig:thirdstage_k3} is the fact that although we change $\lambda$ values in Fig.~\ref{fig:thirdstage_k3}(a) and $\tilde{\lambda}$ values in Fig.~\ref{fig:thirdstage_k3}(b), their effects on the third hop are quite similar. In either case we observe a similar trend, i.e., $k_3$ value decreases as $\lambda$ or $\tilde{\lambda}$ increases.

\section{Appendix: Proof of Lemma \ref{lemma_bound}} \label{section:proof}
In this section, we prove Lemma~\ref{lemma_bound}, first, for the two-hop scenario, then we extend the proof to the $L$-hop case. We use \cite[Thm. 2]{Soysal18} which, for a G/G/1/1 system, requires interarrival times to have DMRL property and service times to have NBUE property. Remember that the updates that depart the service at the first hop, $S_1$ are considered to be the arrivals at the second hop, $R_2$. Then, between a transmitter and a receiver at the second hop, we have interarrivals $R_2 = \sum_{i=1}^{M_1} (X_{k_1:n})_i$ and service times $\tilde{X}_{k_2:n}$. In order to employ \cite[Thm. 2]{Soysal18}, we need to show that $R_2$ has DMRL property and $\tilde{X}_{k_2:n}$ has NBUE property. It is sufficient to show that both $R_2$ and $\tilde{X}_{k_2:n}$ have log-concave density, since log-concavity implies both DMRL and NBUE properties \cite{Marshall07}.

It is given in \cite[Thm. 1.C.54]{Shaked07} that the order statistics of i.i.d. log-concave random variables is log-concave as well. Since shifted exponential has a log-concave density, we conclude that the second hop service time $\tilde{X}_{k_2:n}$ has a log-concave density.

In order to show the log-concavity of $R_2$, we use the fact that a function is log-concave if and only if it is a Polya Frequency function of order 2 \cite[Proposition 21.B.8]{Marshall07}, which is denoted by $PF_2$. We know that the geometric random variable $M_1$ and shifted exponential random variables $X_{k_1:n}$ are log-concave, and hence they have  $PF_2$ densities. In addition, we know that $M_1$ is independent of $X_{k_1:n}$. Now, we can apply \cite[Thm. 6]{Karlin60} that states that if  $X_{k_1:n}$ and  $M_1$ have $PF_k$ densities, and $M_1$ is independent of $X_{k_1:n}$, then $R_2 = \sum_{i=1}^{M_1} (X_{k_1:n})_i$ has a $PF_k$ density as well. Since \cite[Thm. 6]{Karlin60} is stated for any $k$, it holds for $k=2$, and we conclude that $R_2$ has a $PF_2$ density, which, in turm, means that $R_2$ is log-concave.

We have proved that interarrivals to the second hop are DMRL and service time of the second hop is NBUE. Hence, using \cite[Thm. 2]{Soysal18} average age of the second hop is upper bounded by a system where interarrivals are exponential with $E[R_2]$. In the third hop, service times are i.i.d. with the service times at the second hop. Therefore, service time of the third hop is NBUE as well. The interarrivals at the third hop are now $R_3 = \sum_{i=1}^{M_2} (X_{k_2:n} + Z_2)_i$. In order to calculate the upper bound in the second hop, we have already assumed that the interarrivals, $R_2$, are exponential which resulted in $Z_2 = R_2$ to be exponential as well. Then, $R_3$ can be shown to be log-concave using similar ideas above and for exponential $Z_2$. This reasoning holds for each hop in an $L$-hop system, which proves the applicability of Lemma~\ref{lemma_bound} to an $L$-hop system as well. 

\section{Conclusions} \label{conc}
We studied the age of information in a multihop multicast network with a single source updating $n^L$ end nodes where $L$ denotes the number of hops. We showed that when the earliest $k_{\ell}$ transmission policy is implemented at each hop $\ell$, the age of information at the end nodes can be upper bounded by a constant that is independent of $n$. We explicitly characterized an upper bound for an $L$-hop multicast network, and then determined the optimal stopping thresholds $k_{\ell}$ for arbitrary shifted exponential link delays. We note that even when the link delays are exponential we find $k^*_1 =1$ and $k^*_{\ell} > 1$ in hops $\ell>1$. This is because even when there is no shift in service, the random waiting time under exogenous arrivals introduces a random shift to the exponential service distribution in hops $\ell>1$.

\bibliographystyle{unsrt}
\bibliography{IEEEabrv,lib}

\begin{thebibliography}{10}

\bibitem{Mainwaring02}
A.~Mainwaring et~al.
\newblock Wireless sensor networks for habitat monitoring.
\newblock In {\em WSNA}, September 2002.

\bibitem{Papadimitratos09}
P.~Papadimitratos et~al.
\newblock Vehicular communication systems: Enabling technologies, applications,
  and future outlook on intelligent transportation.
\newblock {\em IEEE Communications Magazine}, 47(11):84--95, November 2009.

\bibitem{Kaul12a}
S.~K. Kaul, R.~D. Yates, and M.~Gruteser.
\newblock Real-time status: How often should one update?
\newblock In {\em IEEE Infocom}, March 2012.

\bibitem{Costa16}
M.~Costa, M.~Codreanu, and A.~Ephremides.
\newblock On the age of information in status update systems with packet
  management.
\newblock {\em IEEE Transactions on Information Theory}, 62(4):1897--1910,
  April 2016.

\bibitem{Yates12}
R.~D. Yates and S.~K. Kaul.
\newblock Real-time status updating: Multiple sources.
\newblock In {\em IEEE ISIT}, July 2012.

\bibitem{Najm17}
E.~Najm, R.~D. Yates, and E.~Soljanin.
\newblock Status updates through {M/G/1/1} queues with {HARQ}.
\newblock In {\em IEEE ISIT}, June 2017.

\bibitem{Soysal18}
A.~Soysal and S.~Ulukus.
\newblock Age of information in {G/G/1/1} systems.
\newblock November 2018.
\newblock Available on arXiv:1805.12586.

\bibitem{Bedewy17a}
A.~M. Bedewy, Y.~Sun, and N.~B. Shroff.
\newblock Age-optimal information updates in multihop networks.
\newblock In {\em IEEE ISIT}, June 2017.

\bibitem{Arafa17b}
A.~Arafa and S.~Ulukus.
\newblock Age minimization in energy harvesting communications:
  Energy-controlled delays.
\newblock In {\em Asilomar Conference}, October 2017.

\bibitem{Arafa17a}
A.~Arafa and S.~Ulukus.
\newblock Age-minimal transmission in energy harvesting two-hop networks.
\newblock In {\em IEEE Globecom}, December 2017.

\bibitem{Yates15}
R.~D. Yates.
\newblock Lazy is timely: Status updates by an energy harvesting source.
\newblock In {\em IEEE ISIT}, June 2015.

\bibitem{Bacinoglu15}
B.~T. Bacinoglu, E.~T. Ceran, and E.~Uysal-Biyikoglu.
\newblock Age of information under energy replenishment constraints.
\newblock In {\em UCSD ITA}, February 2015.

\bibitem{Wu18}
X.~Wu, J.~Yang, and J.~Wu.
\newblock Optimal status update for age of information minimization with an
  energy harvesting source.
\newblock {\em IEEE Transactions on Green Communications and Networking},
  2(1):193--204, March 2018.

\bibitem{Arafa18a}
A.~Arafa, J.~Yang, S.~Ulukus, and H.~V. Poor.
\newblock Age-minimal online policies for energy harvesting sensors with
  incremental battery recharges.
\newblock In {\em UCSD ITA}, February 2018.

\bibitem{Arafa18b}
A.~Arafa, J.~Yang, and S.~Ulukus.
\newblock Age-minimal online policies for energy harvesting sensors with random
  battery recharges.
\newblock In {\em IEEE ICC}, May 2018.

\bibitem{Arafa18d}
A.~Arafa, J.~Yang, S.~Ulukus, and H.~V. Poor.
\newblock Online timely status updates with erasures for energy harvesting
  sensors.
\newblock In {\em Allerton Conference}, October 2018.

\bibitem{Baknina18a}
A.~Baknina and S.~Ulukus.
\newblock Coded status updates in an energy harvesting erasure channel.
\newblock In {\em CISS}, March 2018.

\bibitem{Baknina18b}
A.~Baknina, O.~Ozel, J.~Yang, S.~Ulukus, and A.~Yener.
\newblock Sending information through status updates.
\newblock In {\em IEEE ISIT}, June 2018.

\bibitem{Ioannidis09}
S.~Ioannidis, A.~Chaintreau, and L.~Massoulie.
\newblock Optimal and scalable distribution of content updates over a mobile
  social network.
\newblock In {\em IEEE Infocom}, April 2009.

\bibitem{Zhong17a}
J.~Zhong, E.~Soljanin, and R.~D. Yates.
\newblock Status updates through multicast networks.
\newblock In {\em Allerton Conference}, October 2017.

\bibitem{Marshall07}
A.~W. Marshall and I.~Olkin.
\newblock {\em Life Distributions, Structure of Nonparametric, Semiparametric,
  and Parametric Families}.
\newblock Springer, New York, 2007.

\bibitem{Buyukates18}
B.~Buyukates, A.~Soysal, and S.~Ulukus.
\newblock Age of information in two-hop multicast networks.
\newblock In {\em Asilomar Conference}, October 2018.

\bibitem{Shaked07}
M.~Shaked and J.~G. Shanthikumar.
\newblock {\em Stochastic Orders and Their Applications}.
\newblock Springer-Verlag New York, 2007.

\bibitem{Karlin60}
S.~Karlin and F.~Proschan.
\newblock Polya type distributions of convolutions.
\newblock {\em The Annals of Mathematical Statistics}, 31(3):721--736,
  September 1960.

\end{thebibliography}

\end{document}